\documentclass[12pt]{article}
\usepackage{amsmath,amssymb,amsthm,color}
\usepackage{graphicx}
\usepackage{cite}
\usepackage{epsfig}
\renewcommand{\baselinestretch}{2}
\begin{document}
%
\title{Dimension- and lattice-diversified Coulomb excitations in  3D, 2D, 1D- nanotube electron gases, graphene and carbon nanotube}
\author{
\small Cheng-Hsueh Yang$^{*a}$, Chiun-Yan Lin$^{a}$, Chih-Wei Chiu$^{b}$,Chang-Ting Liu$^{b}$, Ming-Fa Lin$^{*a}$ $$\\
\small  $^a$Department of Physics, National Cheng Kung University, Tainan 701, Taiwan\\
\small  $^b$Department of Physics, National Kaohsiung Normal University, Kaohsiung 824, Taiwan.\\
 }
\renewcommand{\baselinestretch}{1}
\maketitle

\renewcommand{\baselinestretch}{1.4}
\begin{abstract}

Elementary electronic excitations, which are due to the Coulomb-field scatterings, present the diverse phenomena in 3D, 2D, 1D-nanotube electron gases, graphene and carbon nanotubes. The critical mechanisms cover the dimension-dependent bare Coulomb potentials, energy dispersions, and free/valence carrier density. They are responsible for the main features, the available excitation channels (the electron-hole regions), the joint van Hove singularities, the undamped/damped collective excitations at small/sufficiently high transferred momenta, the momentum dependences of plasmon frequencies (acoustic and optical modes), and their categories (the intraband and inter-$pi$-band plasmons). There exists certain significant similarities and difference among various systems.  The (momentum/ angular momentum, frequency)-excitation phase diagrams are directly reflected in the propagation of plasma waves.

\vskip 1.0 truecm
\par\noindent

\par\noindent  * Corresponding author. {~ Tel:~ +886-6-2757575-65272.}\\~{{\it E-mail address}: mflin@mail.ncku.edu.tw}
\end{abstract}

\pagebreak
\renewcommand{\baselinestretch}{2}
\newpage

{\bf 1. Introduction}
\vskip 0.3 truecm

The transient screening response in real space and time will become one of the main-stream topics\cite{Nature557;530 , Nature487;82} because of the much progress in the femtosecond optical spectroscopies \cite{LaserPhoton.Rev.4;349 , Spectrochim.Acta B55;31771}. Up to now, there are only few theoretical predictions on the dynamic charge oscillation phenomena. Apparently, the rich relations between the propagating plasma waves and the essential physical properties are required to be clarified thoroughly. For example, the time-dependent charge oscillations of 3D electron gas\cite{Phys.Rev.B38;1647}, being perturbed by a suddenly created point charge, are evaluated within the linear  random-phase approximation. Furthermore, this way allows the decomposition of the charge response into a part due to plasmon modes and another associated with electron-hole pairs (the clear classification of the collective and single-particle excitations). The calculated results, the position- and time-dependent screened charge density distributions, might be suitable for the core-hole creation in simple metals in the limit of large photoelectron energies. A concise picture mainly arises from an electronic "shock wave" which propagates outward from the core hole with some dispersion and the group velocities are centered at the Fermi velocity. Behind this shock wave consists of the static distribution plus a small ringing related to small-transferred-momentum plasmons. Such predictions could be examined in the light of experimental evidences for transient effects on core-hole decay. The similar researches have been conducted on a semi-infinite metal surface\cite{Phys.Rev.B15;3060} and 2D electron gas\cite{Phys.Rev.Lett.18;546}, in which the characteristic times of the transient charge oscillations is about ${0.1-1.0}$ femtosecond. The plasma waves might propagate too quickly to be examined by the ultrafast optical with the super-high resolution of ${\sim\,50-10}$ fs.


Any charge screening phenomena, which come from valence and conduction carriers in any condensed-matter systems can be well characterized by the dimensionless dielectric function or the energy loss function. The latter, corresponding to the emergent layered materials \cite{J.Phys.Soc.Jpn.69;3781 , PhysicaE11;356} and coaxial nanotubes\cite{Phys.Rev.B53;10225}, are successfully defined in
\begin{equation}
Im[\frac{-1}{\epsilon (q,L,\omega )}]\equiv \frac{\sum _{l}Im[-V_{ll}^{eff}(q,L,\omega )]}{\sum _{l,m}V_{lm}(q,L,\omega )/N}
\text{ .} \tag{1.1}
\end{equation}
\\using the developed theoretical framework . The screened response function is a universal dimensionless physical quantity and thus represents the screening ability of the total charges. That is, the comparisons among the different condensed-mater systems are available through this reliable function. When the energy loss spectrum is investigated by its Fourier transform about the transferred momenta and frequencies, the propagating plasma waves, being accompanied with the transport of energy density, are clearly revealed in the real space and time domain . Whether the energy loss function is better/ worse than the screened charge density in describing the dynamic Coulomb-field phenomena is worthy of the systematic investigations according to the consistence of theoretical and experimental viewpoints. The 1D and 2D low-dimensional systems are outstanding candidates in clearly illustrating the diverse carrier dynamics.


The intermediate plasma wave and electron-hole pair excitations, which strongly depend on position and time, are expected to be greatly diversified by the various band properties near the Fermi level. That is to say, the transient charge screenings are very sensitive to the metallic, semi-metallic, zero-gap, narrow-gap and finite-gap behaviors. The critical mechanisms/ pictures could be achieved in 1D three types of carbon nanotubes\cite{J.Phys.Soc.Jpn.71;1820 , Phys.Rev.B50;17744} and 2D few-layer graphene systems\cite{Phys.Rev.B73;144427 , J.Appl.Phys.103;103109}, being easily modulated by the nanotube radius $\&$ chirality\cite{ACSNano7;2205}, number of nanotube/ layer,  stacking configuration\cite{ACSNano9;6333}, low or high doping\cite{Phys.Rev.B56;1430}, temperature, gate voltage\cite{Phys.Rev.B53;10225}, and magnetic field\cite{J.Phys.Soc.Jpn.66;3294}. Of course, the intralayer $\&$ interlayer orbital hybridizations\cite{NanoLett.15;1135}, the intralayer $\&$ interlayer Coulomb interactions \cite{NanoLett.15;2594}, and the external fields are simultaneously unified in the calculations of screened response functions of the developed theoretical framework.  The calculated results, the dynamic charge density oscillations with the direct energy transport, cover their characteristic lengths, times and evolutions. The main contributions from the single- particle and collective excitations are fully distinguished from each other under the various carrier- density environments. How to examine the diversified charge screening phenomena from the high- resolution pump-probe femtosecond optical spectroscopies \cite{LaserPhoton.Rev.4;349 , Spectrochim.Acta B55;31771} needs to be discussed in detail. For example, the low-frequency plasmon modes, with energies below 10 meV, will appear in an intrinsic monolayer graphene at room temperature \cite{Appl.Phys.Lett.99;243502} and thus show them as the coherent plasma waves in the periodical time over 400 fs. The up-to-date experimental measurements might be able to directly observe them in the time domain\cite{J.InfraredMillim.TerahertzWaves30;1319}.


\vskip 0.6 truecm
\par\noindent
{\bf 2. 3D, 2D and 1D-nanotube electron gases }
\vskip 0.3 truecm

The charge screening phenomena, which are due to free carriers (conduction electrons or valence holes) in condensed-matter systems, are frequently simulated as the dielectric responses of electron gases (EGSs ; \cite{Phys.Rev.B97;205118}). Obviously, there are a lot of theoretical\cite{Phys.Rev.B49;16581 , Jpn.J.Appl.Phys.26;1913 , Phys.Rev.B97;205118} and experimental\cite{Phys.Rev.Lett.60;848 , Adv. Mater.25;3264} studies on this main-stream topic. Since the parabolic energy spectra [Figs. 10.1(a) and 10.1(b)] and plane waves are analytic in the form, their electric polarizations behave so under the random-phase approximation. The momentum- $\&$ frequency-dependent formulas of dielectric functions in 3D/ 2D/ 1D-nanotube EGSs are very useful in exploring the essential physical properties, e.g., the spatial induced charge distributions\cite{IEEETrans.Ind.Electron.23;433}, the total Coulomb potentials between two charges \cite{Z.Phys.304;347}, the pair distribution functions\cite{J.Chem.Phys.84;2336}, the exchange $\&$ correlation energies\cite{Phys.Rev.A50;2138},  the Coulomb decay rates\cite{RSCAdv.10;2337},  the modified effective masses, and the renormalization constants \cite{Phys.Rev.Lett.3;351}.This section is mainly focused on the propagation of charge oscillation with plasmon and electron-hole excitations simultaneously.


Without the crystal potential, the Coulomb matrix element, which are related to the initial and final states
\begin{align}
&\left|\left<k_y+q,J+L,h'\left|e^{iqy}e^{iL\phi}\right|k_y,J,h\right>\right|^2\nonumber\\
&=\frac{1}{4}\left\{1+[q^2+(L/r)^2]\right\}^{-6}\times\,\left|1\pm\frac{H^{*}_{12}(k_y+q,J+L)H^{*}_{12}(k_y,J)}{|H^{*}_{12}(k_y+q,J+L)H^{*}_{12}(k_y,J)|}\right|^2.
\tag{2.1}
\end{align},
\\is just equal one for the plane waves. At zero temperature, the real- and imaginary-part dielectric functions can be directly solved within the random-phase approximation. In general there are two technical ways in the formula  calculation: (1) evaluating the Matsubara’s electric polarization function and then doing the analytic continuity for the complex frequency, and (2) deriving ${Im[\epsilon\,]}$ and using the Kramers-Kronig relations in getting ${Re[\epsilon\,]}$. The momentum- and frequency-dependences for  3D \cite{Phys.Rev.B97;205118}, 2D \cite{Phys.Rev.B97;205118 , Phys.Rev.Lett.18;546} and 1D-nanotube \cite{Phys.Rev.B47;6617 , Phys.Rev.B48;5567 , Chin.J.Phys.32;879} electron gases are, respectively, characterized by

\begin{equation}
Re[\epsilon^{3D} (q,\omega)]=1+\frac{1}{2}\frac{k_{TF}^{2}}{q^{2}}+\frac{1}{4}\frac{k_{TF}^{2}}{q^{2}}\frac{k_{F}}{q}[(1-x_{1}^{2})ln\left | \frac{x_{1}+1}{x_{1}-1} \right |+(1-x_{2}^{2})ln\left | \frac{x_{2}+1}{x_{2}-1} \right |]
\text{ ,} \tag{2.2a}
\end{equation}

\begin{equation}
Im[\epsilon^{3D} (q,\omega)]=\left\{\begin{matrix}
\frac{\pi}{4}\frac{k_{TF}^{2}}{q^{2}}\frac{k_{F}}{q}\frac{\hbar\omega }{E_{F}} \qquad \omega \leq \frac{\hbar}{2m}(-q^{2}+2qk_{F})\\
\frac{\pi}{4}\frac{k_{TF}^{2}}{q^{2}}\frac{k_{F}}{q}[1-(\frac{q}{2k_{F}}-\frac{m\omega }{\hbar qk_{F}})] \qquad \omega \geq  \frac{\hbar}{2m}(-q^{2}+2qk_{F})\\
0 \qquad others.
\end{matrix}\right.
\tag{2.2b}
\end{equation}

The discussion regarding kinematics and the value of k is the same as in the 3D case. Hence , the dielectric of 2D case can be expressed by

\begin{equation}
\epsilon (q,\omega )=\epsilon _{0}-\frac{2\pi e^{2}}{q}\chi (q,\omega )
\text{ ,} \tag{2.3}
\end{equation}

\begin{equation}
\begin{split}
Re[\chi ^{2D}(Q,\omega )]&=-\frac{k_{F}^{2}}{2\pi ^{2}E_{F}}[-1-sgn(\omega -Q^{2})\Theta[\left | \omega -Q^{2} \right |/(2Q)]-1]\frac{1}{Q}\sqrt{\left | \frac{\omega -Q^{2}}{2Q} \right |^{2}-1} \\
&+sgn(\omega +Q^{2})\Theta[\left | \omega +Q^{2} \right |/(2Q)]-1]\frac{1}{Q}\sqrt{\left | \frac{\omega +Q^{2}}{2Q} \right |^{2}-1}]
\text{ ,}
\end{split}
 \tag{2.4a}
\end{equation}

\begin{equation}
\begin{split}
Im[\chi _{a}^{2D}](Q,\omega )&=\frac{k_{F}^{2}}{2\pi E_{F}}2\int_{k_{a}}^{1}\frac{kdk}{\sqrt{4Q^{2}k^{2}-(\omega -Q^{2})^{2}}} \\
&=-\frac{k_{F}^{2}}{2\pi E_{F}}\frac{1}{Q}\sqrt{1-\frac{1}{4}(\frac{\omega }{Q}-Q)^{2}}-\sqrt{(1-\frac{Q^{2}}{4})-\frac{\omega }{2}-\frac{\omega ^{2}}{4Q^{2}}}
\text{ ,}
\end{split}
\tag{2.4b}
\end{equation}

\begin{equation}
\begin{split}
Im[\chi _{b}^{2D}](Q,\omega )&=\frac{k_{F}^{2}}{2\pi E_{F}}2\int_{k_{b}}^{1}\frac{kdk}{\sqrt{4Q^{2}k^{2}-(\omega -Q^{2})^{2}}} \\
&=-\frac{k_{F}^{2}}{2\pi E_{F}}\frac{1}{Q}\sqrt{1-\frac{1}{4}(\frac{\omega }{Q}-Q)^{2}}
\text{ .}
\end{split}
\tag{2.4c}
\end{equation}

\begin{equation}
Im[\chi ^{2D}](Q< 2, 0\leq \omega < -Q^{2}+2Q)=Im[\chi _{a}^{2D}(Q,\omega )]
\text{ ,} \tag{2.5a}
\end{equation}

\begin{equation}
\begin{split}
Im[\chi ^{2D}](Q\geq 2 , Q^{2}-2Q\leq \omega \leq  Q^{2}+2Q)&=Im[\chi ^{2D}](Q< 2 , -Q^{2}+2Q\leq \omega \leq  Q^{2}+2Q) \\
&=Im[\chi _{b}^{2D}(Q,\omega )]
\text{ .}
\end{split}
\tag{2.5b}
\end{equation}
\\where Q=q/2${k_F}$, and ${\epsilon_0}$=1 is background dielectric constant . In the 1D-nanotube case which is equivalent to RPA is expressed by

\begin{equation}
\epsilon (q,J,\omega )=\epsilon _{0}-V(q,J;r)\chi (q,L,\omega )
\text{ ,} \tag{2.6}
\end{equation}

where ${\epsilon_0}$=2.4 is the background dielectric constant ,and ${V(q,L;r)=4 \pi e^2 I_{L}(qr) K_{L}(qr) }$ is the Coulomb interaction for electrons on the tubule. ${\chi(q,L,\omega)}$ is the response function (the RPA bubbles):

\begin{equation}
\chi (q,L,\omega )\equiv \sum_{occ}^{J}\chi ^{J}(q,L,\omega )
\text{ ,} \tag{2.7a}
\end{equation}

\begin{equation}
Re\chi^{J}(q,L,\omega )=\frac{m^{*}}{2\pi q}ln\left | \frac{\omega ^{2}-E_{-}^{2}(q,J,L)}{\omega ^{2}-E_{+}^{2}(q,J,L)} \right |
\text{ ,} \tag{2.7b}
\end{equation}

and

\begin{equation}
Im\chi ^{J}(q,J,\omega )=\left\{\begin{matrix}
\frac{m^{*}}{2\pi ^{2}q} \qquad if \left | E_{-}(q,J,L) \right |< \omega < \left | E_{+}(q,J,L) \right |\\
\frac{-m^{*}}{2\pi ^{2}q} \qquad  if \left | E_{+}(q,J,L) \right |< \omega < \left | E_{-}(q,J,L) \right |\\
0 \qquad  otherwise .
\end{matrix}\right.
\text{ ,} \tag{2.7c}
\end{equation}


The above-mentioned equations present the different conditions for the non-zero ${Im[\epsilon\,]}$, since the conservation of momentum/angular momentum and energy, the Pauli exclusion principle and The Fermi-Dirac distribution need to be satisfied during the electron-electron Coulomb interactions. They will determine the available electron-hole pair excitation regions in momentum-frequency space, with the Fermi-momentum-related boundaries. It should be noticed that a hollow cylinder has a lot of 1D energy subbands with the well-behaved standing waves, so it displays the decoupled Coulomb excitation modes of distinct angular-momentum transfers. It should be noticed that the non-physical results might appear at very low carrier densities. This drawback is induced by the approximate excitation model. Apparently, the different dimensions play a critical role in electronic properties and electron-electron Coulomb interactions simultaneously, leading to the diversified charge screening behaviors.


The real- and imaginary-part dielectric function, which mainly comes from the available single-particle transition channels, present the dimension-enriched strong responses. Figures 10.2(a)-10.2(p) clearly illustrate the prominent structures at the specific frequencies under the fixed momentum transfer or Fermi energy. After a very slight broadening, ${Im[\epsilon\,]}$ of the 3D system in Fig. 10.2(b) must be vanishing for a specific momentum at the excitation frequency more than ${E_q+v_fq}$ [the Fermi velocity ${v_F=k_F/m}$], i.e., the available single-particle excitation channels are thoroughly absent through the higher frequency beyond the highest electron-hole pair boundary [Figs. 10.4(a)-10.4(f)]. This behavior generates a cusp structure there, being also revealed in 2D and 1D electron gases [discussed later in Figs. 10.2(f), 10.2(h), 10.2(j) and 10.2(l). Moreover, ${Im[\epsilon\,]}$ shows a modified left-side shoulder or cusp structure, as displayed in Fig. 10.2(b) at the distinct Fermi levels. Its corresponding frequency gradually deviates from the maximum electron-hole excitation frequency in the increment of $E_F$, where the strength of single-particle excitations decline simultaneously. The second special structure at the lower frequency should come from the joint van Hove singularities closely related to the Fermi-momentum e initial or final states during the electron-electron scatterings. On the other side, ${Re[\epsilon]}$ in Fig. 10.1(a)  displays the opposite structure, as well as the drastic variations between the positive and negative values. The real-part response function approaches to zero at the higher frequency, where ${Im[\epsilon\,]}$ also vanishes before the creation of Landau dampings. Apparently, this will generate the prominent resonance phenomenon of free-carrier oscillations, and the plasmon intensity is determined by the first derivative of ${Re[\epsilon\,]}$ versus $\omega$ at a zero point \cite{Phys.Rev.B34;979}. Such feature is greatly enhanced by the higher Fermi levels and the large Fermi momenta. In addition to the strong Fermi-energy dependences, the dialectic function is very sensitive to the change of momentum transfer, in which the screening ability declines with the increasing $q$’s. At very small $q$’s ${q\rightarrow\,0}$, the first and second cusp structures would almost merge together.  While the transferred momentum is sufficiently high, ${Im[\epsilon\,]}$/${Re[\epsilon\,]}$ only displays a broad peak/a monotonous decrease from the positive value to zero  [the green curve in Fig. 10.2(d)/10.2(c)]. The monotonous $E_F$- and $q$-dependences are deduced to come from the fully isotropic energy spectrum and ideal plane-wave functions.


The prominent screening behaviors of conduction electrons, which are created by the unusual Fermi surfaces, are also revealed in other dimensional electron gases. However, the detailed structures of dielectric functions are totally distinct among 2D, 1D-rubulw and 3D electron gases. ${Im[\epsilon\,]}$ [${Re[\epsilon\,]}$] of 2D electron gases, as shown in Fig. 10.2(f) [Fig. 10.2(e)] exhibits a cusp and a modified shoulder  [two strongly modified peaks with the positive and negative values] at the lower and higher frequencies, respectively. Also noticed that the latter would become a straight shoulder form at very small momentum transfers, being estimated from Fig. 10.2(h). Its frequency is close to the highest electron-hole excitations [the upper boundary in Figs. 10.4(g)-10.4(l)]. This composite structure sharply contrasts to a specific cusp of the 3D case [Figs. 10.2(b) and 10.2(d)]. Such result means the existence of three joint van Hove singularities in the Fermi-momentum-related Coulomb scatterings. As for the 1D-nanotube electron gas, the five occupied conduction subbands [the ${J=0}$, ${\pm\,1}$ and ${\pm\,2}$ in Fig.10.1(b)] could exhibit the multi-channel Coulomb excitations under each $L$-decoupled mode, e.g., the ${L=0}$ [Figs. 10.2(i)-10.2(l)] and ${L=1}$ [Figs. 10.2(m)-10.2(p)] dynamic transitions. While the transferred is enough large, e.g., ${q\ge\,0.5}$ $k_F$ [the purple and green curves in Fig. 10.2 (l)], the imaginary-part dielectric function clearly display the obvious   plateau structures within the specific frequency ranges, being confined by the different electron-hole excitation boundaries [discussed later in Fig. 10.4(o)]. Their number and widths grow as the transferred momentum increase [Fig. 10.2(l)]; furthermore, these features strongly depend on the Fermi energy [Fig. 10.2(l)]. In addition, plateaus are replaced by prominent symmetric peaks under the enough narrow case. Apparently, a plateau [a pair of shoulders]/a symmetric peak in ${Im[\epsilon]}$ corresponds to two logarithmic peaks/the positive and negative asymmetric peaks in ${Re[\epsilon]}$ [Figs. 10.2(i) and 10.2(k)]. The real-part dielectric functions display more special structures in the low-dimension systems, especially for the 1D ones.Moreover, each ${Re[\epsilon]}$ would gradually vanish at the higher frequency without the Landau damping below the critical momentum, since it presents the collective oscillations of the whole free carriers.


The elementary excitations in 1D-nanotube systems is very useful in fully exploring the fundament screening behaviors, such as, the exact resonance frequencies of coherent charge oscillations and the $f$-sum rule [details in Eq. (7) in \cite{Phys.Rev.B47;6617 , Benjamin,NewYork}]. In general, there are two methods in searching a plasmon frequency: (1) using a specific zero point of the real-part dielectric function under the small/vanishing imaginary-part one, and (2) a prominent structure in energy loss spectrum [in the whole book]. The former is very suitable for condensed-matter systems with simple energy bands and wave functions, such as, the low-doping semiconductors of 2D quantum wells \cite{NatureCommunications9;2254} and 1D quantum wires \cite{PhysicaELowDimens.Syst.Nanostruct.79;20}. However, this technique might become meaningless in the absence of zero points, the Landau damings are very prominent through a lot of valence electrons \cite{Phys.Rev.B34;979}. For example, the graphene-related materials possess the high-density $\pi$-valence electrons and frequently induce the non-zero ${Re[\epsilon\,]}$ at the plasmon frequencies.  The previous study shows that an analytical formula of dielectric function [Eq. (2.4)] could be achieved for the1D-tubule electron gas\cite{Phys.Rev.B47;6617}. Corresponding to a strong single-particle response, ${Re[\epsilon\,]}$ must present a drastic change between the positive and negative values. As a result, there exist two zero points, in which the lower- and higher-frequency ones indicate the very strong and weak suppressions of electron-hole pair excitations. The specific real-part zero point is capable of providing a resonant phenomenon and thus determines the collective excitation frequency, being thoroughly examined for any momentum- and angular-momentum-decoupled modes [e.g., Figs. 2 and 4 in \cite{Phys.Rev.B47;6617}]. Specifically, we can utilize the ${Im[-1/\epsilon\,]}$-related $f$-sum to define the oscillator strength of collective and single-particle Coulomb excitations. And then, both of them are, respectively, evaluated from the first derivative of ${Re[\epsilon\,]}$ versus ${\omega}$ and the frequency integration for ${\omega\,Im[-1/\epsilon\,]}$  [details in Eqs. (9) and (10) of \cite{Phys.Rev.B47;6617}]. For each $L$-decoupled avaulable channel, there might have the multi-plasmon modes and several electron-hole excitation regions. The detailed examinations clearly identify the conservation of charge number and further illustrate the critical factors, the Coulomb interactions and carrier density, in dominating the rich screening phenomena.


Apparently, the energy loss spectra strongly depend on the Fermi energy and momentum transfer, as revealed in Fig. 10.3(a) and 10.3(h). When the momentum transfer is much smaller the Fermi momentum, e. g., ${q=0.1}$ $k_F$ in Fig. 10.3(a), the 3D electron gas presents a very strong response using a universal quantity of ${>500}$ for any Fermi levels. Furthermore, both frequency and height of plasmon peak grow in the increase of $E_F$. These features only directly more free conduction electrons in supporting the coherent charge oscillations under the absence of Landau Landau dampings. The plasmon frequency is also enhanced by the larger momenta [Fig. 10.3(b)]. However, the strength of collective excitations shows the opposite behavior. This indicates a better coherence through the long-wavelength carrier vibrations. With the transferred momentum beyond $k_F$, the plasmon structure is seriously suppressed by the single-particle excitations and even might be replaced by the latter [discussed later in Fig. 10.4(c)]. There also exits a prominent peak in each 2D-/1D-case [Figs. 10.3(c)-10.3(f)], corresponding to the coherent oscillation of all free conduction electrons. Apparently, their main features, strength and frequency, are greatly diversified by the dimensionality and the cylindrical symmetry. For example, the ${L=0}$ and ${L=1}$ modes in a 1D-nnanotube gas are quite different from each other [Figs.10.3(e)-10.3(f) and Figs.10.3(g)-10.4(h)], since they originate from the achiral and chiral charge screenings, respectively.


The screened response functions in the momentum-frequency domain [Figs. 10.4(a)-10.4(x)], which directly indicate the intrinsic ability of charge screenings, are able to provide the full information about Coulomb excitations and the other dynamic behaviors. The 3D electron gases, as shown in Figs. 10.4(a)10.4(f) under the various Fermi energies, exhibit the single-particle excitations within the range ${\hbar^2\,q^2/2m}$+${qk_F/m\le\,omega\le}$${\hbar^2\,q^2/2m}$-${qk_F/m}$, according the conservation of momentum and energy from the ${\bf k_F}$-state-related Coulomb scatterings [Eq. 10.1]. Only this continuous ${[q, \omega\,]}$-region could survive and the other ones are thoroughly absent, mainly owing to the unique intraband channel of conduction electrons. Such behavior further illustrates the significance of the isotropic Fermi/spherical surface . In the absence of  single-particle excitations, a prominent plasmon peak is obviously revealed at small $q$,s and even beyond the Fermi momenta. The oscillation frequency of coherent charge screenings is finite at the long wavelength limit, thus, belonging to an optical mode \cite{PhysicaELowDimens.Syst.Nanostruct.54;267}.
It is mainly determined by three critical factors: (1) the momentum-dependent bare Coulomb potential [${4\pi\,e^2/q^2}$], (b) minimum excitation energy [${q^2/2m}$], and (3) free carrier density , being also suitable for 2D and 1D \cite{J.Appl.Phys.116;024309} conduction electrons/valence holes.  The approximate result from the specific zero point of ${Re[\epsilon]}$ under RPA is ${\omega\,\omega_0}$${+O(q^2)}$ [${\omega_0^2\,=4\pi\,ne^2/m}$, $n$  carrier density;\cite{Phys.Rev.B34;979}].  Its value is very sensitive to the changes   of free carrier densities, or ${\omega_p\,(q=0)/E_F}$ decreases for the increase of the Fermi energy. When the transferred momentum  is enough large [$q_{c1}$], such plasmon will experience the Landau damping. Its strength quickly declines in the increment of $q$ and then vanishes at the higher $q_{c2}$. Both critical momenta are very close too each other. Specifically, ${q_c/k_F}$ is larger under  the lower Fermi energy. The undamped plasmon modes are relatively observed under the smaller $E_F$-case. It should be noticed that the dispersion relation for plasmon frequency with momentum could be approximately obtained from the zero point of the real-part dielectric function by using the small-$q$ and large-$\omega$ conditions. The above-mentioned electron-hole excitation region are also found in 2D\cite{Phys.Rev.B97;205118} and 1D electron gases\cite{Phys.Rev.Lett.77;135}. However, the momentum-dependent coherent charge oscillations are drastically changed by the dimension-dominated Coulomb potential. According to three significant mechanisms,  their frequency exhibits the ${\sqrt q}$-dependence at the long wavelength limit. This is the well-known 2D acoustic plasmon, as verified from the doped semiconductor quantum wells\cite{NatureCommunications9;2254 , PhysicaELowDimens.Syst.Nanostruct.79;20}.


The 1D-nanotube systems, as clearly illustrated in Figs. 10.4(m)-10.4(r) and 10.4(s)-10.4(x), respectively, for the ${L=0}$ and  ${L=1}$ modes, present the rich and unique Coulomb-excitation phase diagrams. The critical mechanisms cover the dimensionality and cylindrical symmetry. The occupied conduction energy subbands, which possess two discrete Fermi-momentum states/one ${k_y=0}$  band0edge state in a separate energy spectrum [Figs. 10.1(b)], can create certain electron-hole excitation boundaries (details in \cite{Phys.Rev.B47;6617}). Their excitation frequencies of the ${L}$-decoupled mode are expressed as

\begin{equation}
E_{\pm }(q,J,L)=\frac{q^{2}\pm 2k_{F}(J)q}{2m^{*}}+\frac{L^{2}+2JL}{2m^{*}r_{1}^{2}}
\text{ .} \tag{2.6}
\end{equation}


Any single-particle and collective Coulomb excitations are thoroughly absent within the vacant regions, since the serious limit due to the momentum and energy conservations only presents in the 1D metallic systems. Apparently, the screened response functions are greatly by the achiral and chiral charge oscillations. The $L$-dependent    plasmon modes under a fixed Fermi level are totally different from one another. For example, at ${E_F=0.5}$ eV [Figs. 10.4(o) and 10.4(u)], the five occupied conduction subbands [Fig. 10.1(b)] are able to generate three-${L=0}$ and five-${L=1}$ plasmon modes, respectively [part of them too weak to observe; of details in \cite{Phys.Rev.B47;6617}]. Furthermore, the highest-frequency plasmon will dominate the whole oscillator strength, i.e., this mode should play the most important roles in determining the other essential properties, e.g., the quasiparticle lifetimes. Very interesting, under the long wave length, the ${L=0}$ plasmon presents the ${q|ln(qr)|^{1/2}}$ dependence, being similar to that in an armchair carbon nanotube. Three acoustic plasmons do not experience the Landau dampings at small $q$s, while they decay into the neighboring electron-hole region and even disappear through the opposite case.  The calculated results show that only the highest-frequency mode could be observed from the experimental examinations, where it is induced by the coherent oscillations of the whole free conduction electrons along the nanotube axis. The other two plasmon modes are strongly suppressed by the single-particle excitations of the lowest conduction band. On the other side, the chiral ${L=1}$ modes [Fig. 10.4(u)] is due to the charge vibrations in the longitudinal and transverse directions. As a result, more complicated electron-hole excitation regions/boundaries and plasmon modes come to exist. Five plasmons have finite oscillation frequencies at ${q\rightarrow}$ and thus belong to  the optical modes\cite{PhysicaELowDimens.Syst.Nanostruct.54;267} .


\vskip 0.6 truecm
\par\noindent
{\bf 3. $n$-type graphenes}
\vskip 0.3 truecm

A doped monolayer graphene is totally different from a pristine one in Coulomb excitations, since free electrons/holes can create the intra-$pi$-band transitions between  the same conduction/valence band. In addition to the middle-frequency $\pi$ plasmons ${\omega_p\,>5}$ eV, the lower-frequency intraband ones come to exist under the doping cases. The coherent charge oscillations, which come from all the valence $\pi$ electrons are too quick to be observed through the super-high-resolution experimental resolution (about 50 fs in \cite{Chin.Phys.Lett.29;070601}). The studying focus covers the main features of intra- and inter-$pi$-band electron-hole excitations and the acoustic plasmons.

The rigid-band model, which the Fermi level presents a blue shift, is utilized to characterize the $\pi$-electronic states of a doped monolayer graphene, e.g., the alkali-adsorbed system by the ARPES examinations \cite{Rev.Mod.Phys.75;473}. Figure 10.5(a) clearly shows a pair of valence and conduction bands linearly intersecting at the K/K$^\prime$ point. This Dirac-cone band structure, ${E^{c,v}({\bf k})=\pm\,v_F|{\bf k}|}$ [the Fermi velocity of ${v}F=3\gamma_0\,b/2]$, is isotropic along the different directions, in which its Bloch wave function is the symmetric/ anti-symmetric superposition of the A- and B- sublattice- related tight-binding functions for valence/ conduction states. The analytic formulas of the low-lying electronic states are available in deriving the approximate dielectric function at the small-momentum-transfer limit. The previous study shows that  ${\epsilon}$ hardly depends on the direction of the transferred momentum and then could be used to investigate the quasiparticle properties enriched by the $n$-/$p$-type doings, such as, the $E_F$-dependent energy spectra and lifetimes \cite{RSCAdv.10;2337 , Phys.Rev.B26;4421}.


The main features of a gapless Dirac cone play critical roles in the bare and screened response functions (Figs. 10.6 and 10.7). Its conduction and valence electrons can generate the ${c\rightarrow\,c}$ and ${v\rightarrow\,c}$ Coulomb excitation channels (Fig. 10.5(a)). The maximum momentum-dependent excitation frequencies of the former ${v_Fq}$ are associated with the  Fermi-momentum states, being identical to the minimum one of the latter (details in Fig. 10.8). Obviously, the imaginary-part dielectric functions, which are, respectively, shown under the various Fermi levels and transferred momenta (Figs. 10.6(b) and 10.6(d)), is dominated by a modified asymmetric peak during each electric polarization. Since it appears in the left-hand side, so the intraband single-particle excitations should be the dominating ones.  By using the small-$q$ approximation, the analytic formula of dielectric function, the inversely proportional to ${\sqrt {v_Fq-\omega}}$/ ${\sqrt {\omega\,-v_Fq}}$ by the ${c\rightarrow\,c}$/${v\rightarrow\,c}$ channels, could be achieved for the linear valence and conduction bands \cite{Phys.Rev.B34;979}. This structure experiences more serious modifications at smaller Fermi levels and momentum transfers, e.g., the almost symmetric peaks through the comparable co-existence of two-kind excitation channels at ${q=0.1 k_F}$ $\&$ $E_F$=0.1 eV, and ${E_F=0.5}$ eV $\&$ ${q=0.05}$ $k_F$  (the red curves in Fig. 10.6(b)). The corresponding ${Re[\epsilon\,]}$s exhibit the opposite asymmetric peaks, i. e., they display a drastic change from the positive to divergently negative value near the excitation boundary of ${\omega\,=v_Fq}$. Based on the theoretical analyses, there also exist the lower electron-hole excitation boundaries: ${v_Fq}$ ${-2v_Fk_F}$ and ${2v_Fk_F}$${-v_Fq}$, respectively, due to the intyraband and interband transitions. The available electron-hole pair regions in the whole ${q, \omega\,}$-domain are very important in the further understanding on the undamped/damped plasmon modes and the other dynamic behaviors of charge screenings.


The coherent oscillations of free conduction electrons in doped graphene systems are revealed as a prominent plasmon peak in  each energy loss spectrum. The screened response functions, which  strongly depend on the Fermi levels and momentum transfers, are displayed in Figures 10.7(a) and 10.7(b), respectively. Under the small  $q$s, the frequency and strength of collective excitations grow in the increment of $E_F$ because of more high carrier density. However, at a fixed Fermi level, the momentum dependence is non-monotonous. For example, a $E_F$-0.5-V graphene (Fig. 10.7(b)) presents an enhanced screening ability as $q$ increases from zero. The strongest plasmon peak is achieved and then quickly declines beyond the critical momentum, e.g., the great reduce of energy loss function at ${q=k_F}$ [the green curve], where the inter-$\pi$-band Landau dampings just starts to accompany with it. Specifically,   this pronounced plasmon mode cannot co-exist with the intra-$\pi$-band electron-hole excitations, being totally different from those in 3D and 2D electron gases (Figs. 10.4(a)-10.4(h)). As for few-layer doped graphene systems, more plasmon peaks, being classified into the intra- and inter-conduction-band ones, have been thoroughly investigated in the previous book \cite{CRCPress}. In short, the dimensionless energy loss spectra of metallic layered graphenes are diversified by the layer number, stacking configuration, free carrier density, gate voltage, magnetic field, and temperature.


The $n$-type monolayer graphene presents the rich and unique ${[q, \omega\,]}$-excitation phase diagrams, as clearly indicated in Figs. 10.8(a)-10.8(f) for the various Fermi levels. The similar Coulomb excitations are revealed in $p$-type monolayer systems. Furthermore, the anisotropic feature is negligible, since they are insensitive to the direction of transferred momentum. There are two vacant regions without any electron-hole excitations. The first one, which occurs below the intraband boundary [the red line of ${v_Fq=2E_F}$], is related to the minimum intraband excitation frequency of the Fermi surface. Furthermore, another one is situated between the maximum intraband boundary and the minimum interband excitation energy. Apparently, the undamped collective excitations could survive in the second region, especially for the small-${q}$-plasmon mode due to free carriers. At the long wavelength limit, the approximate plasmon frequency is obtained by evaluating the higher-frequency zero point in the real-part dielectric function. That is, ${\omega_p\propto\sqrt {n_{f}q}}$, being similar to that of the 2D electron gas \cite{Phys.Rev.Lett.18;546}. The main reason lies in the fact that both 2D systems have the identical bare Coulomb potential and momentum dependence of excitation frequency ${\omega_{ex}\propto\,v_Fq}$]. This phenomenon further illustrates the almost similar single-particle excitation behaviors among any metallic 3D, 2D and 1D condensed-matter systems, regardless of their energy dispersions of conduction bands. For example, at zero temperature, the quasiparticle states near the Fermi level, respectively, exhibit three kinds of wave-vector-dependent Coulomb decay rates in the 3D, 2D, and 1D metals through the electron-hole pair excitations: ${|k-k_F|^2}$, ${|k-k_F|^2}$${ln|k-k_F|}$, and ${|k-k_F|}$ \cite{Phys.Rev.B26;4421}. Similarly, the Fermi-momentum states show the temperature dependences, $T^2$, -${T^2lnT}$ and $T$, respectively, for 3D, 2D and 1D systems \cite{J.Phys.Soc.Jpn.69;607 , J.Phys.Soc.Jpn.69;3429}. Also, the free carrier density plays a dominating role on the main features of conduction-band plasmon. The EELS measurements \cite{New York and London: Plenum370 , Phys.Rev.Lett.80;4729} are relatively easily to examine the coherent charge oscillations in the higher-$E_F$ systems, e.g., the highest frequency, intensity and critical momentum at ${E_F}$=1.5 eV among six doping cases. It should be noticed that doped monolayer graphenes do not show the discrete inter-$\pi$ band plasmon modes, as revealed in doped carbon nanotubes (discussed later in Figs. 10.11(c)-10.11(f).). However, there exists a very strong ${5-7}$-eV $\pi$ plasmon mode in any graphene-related system with a well-behaved carbon-${sp^2}$ bondings.


\vskip 0.6 truecm
\par\noindent
{\bf 4. doped carbon nanotubes }
\vskip 0.3 truecm

For doped carbon nanotubes , there exist conduction electrons and high-density $\pi$-valence carriers, thus leading to the intra- and inter-$\pi$-band excitations simultaneously. This makes them quite different 1D-nanotube electron gas. Furthermore, the elementary excitations and the propagating plasma waves in five typical systems deserve a closer examination, especially for the diversified phenomena. The up-to-date experimental verifications also discussed in detail.


Specifically, only Coulomb excitations of the ${L=0}$ mode in the ${[18, 0]}$ carbon nanotube is chosen for a model investigation. A pristine systems has a narrow gap of ${E_g\sim\,50}$ eV (inset in Fig. 10.5(b)).After the $n$-type doping, the Fermi level is assumed to be rigidly shifted to the conduction subbands. For example, ${E_F}$=0.50, 0.75 and 1.0 eVs, respectively, correspond to  two [${J^c}$=12 $\&$ 24], four [12, 13, 23$\&$ 24] , and six occupied ones [11, 12, 13, 23, 24$\&$ 25].  The energy dispersions are parabolic near the band-edge states, but quickly become linear as $k_y$ increases from zero. Under the vanishing angular-momentum transfer, the non-vertical Coulomb excitations come to exist through the ${J^c\rightarrow\,J^c}$ and ${J^v\rightarrow\,J^c}$ transition channels. The latter, which will play a critical role in Landau damping, is non-negligible because of many valence electrons.


The single-particle excitations, as clearly illustrated in Figs. 10.9(a)-10.9(d), directly the main features of electronic energy spectra and wave functions. For a doped ${[18, 0]}$ carbon nanotube with ${E_F=0.5}$ eV, ${Re[\epsilon]}$  only presents a prominent peak with a minor modification on the symmetric structure under the various momentum transfers (Fig. 10.9(b)). Both strength and frequency grows with the increasing momentum. This unusual structure   mainly comes from the intraband excitations within the ${J^c=12}$/24 conduction subbands. Furthermore, their composite energy dispersions near the Fermi level, as discussed earlier in Chap. 4, play the critical roles in determining the joint van Hove singularity. And then, ${Re[\epsilon]}$ exhibits one pair of asymmetric peaks in the drastic change of positive and negative values (Fig. 10.9(a)). Apparently, the dielectric function is strongly affected by the Fermi level. There might have more pronounced charge responses at the higher $E_F$’s, such as, two and three symmetric peaks/pairs of anti-symmetric peaks, respectively, under ${E_F=0.75}$ and 1.0 eVs [the orange and green curves in Fig. 10.9(d) and 10.9(c)]. Under the normal cases, the real-part dielectric function approaches to disappear at the higher frequency without any electron-hole excitations [${Im[\epsilon\,]=0}$], indicating the existence of the whole free carrier oscillations. In addition, the inter-$\pi$-band electron-hole excitations, which will show the right-hand asymmetric peaks in the square-root form, are thoroughly absent within the lower-frequency range. For example, these channels show the divergent peaks at ${\omega\,>2E_F}$ under ${q=0.1}$ $k_F$ and ${E_F=0.5}$ eV   [the green curve in the inset of Fig. 10.9(b); also discussed later Figs. 10.11(a)-10.11(d)].


Each energy loss spectrum of the ${L=0}$ mode shows a prominent plasmon peak under any transferred momenta [Fig. 10.10(a)] and free carrier densities [Fig. 10.10(b)]. Apparently, this prominent screened response mainly originate from the coherent charge oscillation of conduction electrons in the degenerate ${J^c=12}$ and 24 subbbands. Without the Landau dampings, both intensity and frequency obviously grow in the increment of $q$/$E_F$. However, its strength declines quickly, when the transferred momentum is close to or beyond the critical one, e.g., ${Im[-1/\epsilon\,]}$ at ${q=k_F}$ and ${E_F=0.5}$ eV [the green curve in Fig. 10.10(a)]. The inter-$\pi$-band electron-hole excitations are responsible for the Landau damping, being in sharp contrast with the main mechanisms [the intraband ones] in the 3D/2D/1D-naotube electron gases [Figs. 10.4(a)-10.4(r)]. Also, the screened response functions are different from those [Figs. 10.7(a) and 10.7(b)] of doped monolayer graphenes in terms of the plasmon-mode frequency and intensity, mainly owing to the dimensionalities and energy dispersions.


In addition to the ${L=0}$ mode, the other ${L}$-dependent elementary excitations, which corresponding to the non-vertical Coulomb excitations of the neighboring subbands, are clearly revealed in any carbon nanotubes . Under the $n$-type doping, they could be further classified into the interband ${c\rightarrow\,c}$ and ${c\rightarrow\,c}$ channels [arrows in Fig. 10.5(b)], in which the second mechanisms are absent in 1D-nanotube electron gas. The special structures of the ${L=1}$ real- and imaginary-part dielectric functions [Figs. 10.9(e)-10.9(h)] are similar to those of the ${L=0}$ ones [Figs. 10.9(a)-10.9(f)], since they mainly come from the Fermi-momentum and band-edge states, with the unusual joint density of states. Apparently, their main features, the number, intensity and frequency of the van Hove singularities, are quite different among the $L$-decoupled single-particle excitations. In addition, only the lower-frequency former are displayed in these figures. As a result, a prominent plasmon peak in each energy loss spectrum [Fig. 10.10(c) and (d)] just corresponds to the coherent charge oscillation of the whole free conduction electrons. It might  experience the undamped or damped charge screenigs, depending on the magnitude of momentum transfer. Furthermore, the interband electron-hole excitations serve as the critical role of Landau daming.


The Coulomb excitation phase diagrams of the doped [18, 0] carbon nanotubes, which are characterized in the [${q, L=0, \omega\,]}$-domains under the various   conduction-electron densities, present the diverse phenomena. Figures 10.11(a) and 10.11(f) clearly display more electron-hole pair regions, compared with those in metallic monolayer graphenes [Fig. 10.8(a)-10.8(f)]. The highest intraband boundary and the lowest interband one are split by the original energy gap. More single-particle excitation channels are generated through the increasing Fermi energy. For example, the ${E_F-0.75}$/1.0-eV systems, with the degenerately occupied four/six conduction subbands [Fig 10.5(b)], show the complicated excitation behaviors due to the distinct intraband channels. The most important intranband plasmon mode, corresponding to the strongest peak in each energy loss spectrum, should be associated with the coherent oscillations in the whole occupied conduction subbands even under the very high carrier densities [e.g., the ${E_F}$-1.5 eV carbon nanotube in Fig. 10.11(f)]. Its frequency approaches to the ${q\sqrt {ln(qr)}}$ dependence at the long wave-length limit, being determined by the barge Coulomb potential., excitation frequency and carrier density. The approximate result could be obtained from the vanishing real-part dielectric function through the limits of large $\omega$s and small $q$s \cite{Phys.Rev.B53;15493}. Such collective excitations reveal the rather strong responses before the critical moment transfer and their spectral intensities decline rapidly inside the specific inter-$\pi$-band electron-hole region, but not the intraband one. Very interesting, whether there  exist one or two acoustic plasmon modes is determined by the free carrier density [the number of occupied conduction subbands]. For example, the low-density conductions electrons, corresponding to the second conduction subbands for the ${E_F}$=0.7 eV/1.0 eV system , present the weak acoustic plasmon. Moreover, the different  inter-$\pi$-band plasmons, with the weak, but significant peaks in EELS, come to exist at the higher frequencies . Such optical modes thoroughly disappear in any graphene systems [Figs. 10. 8(a)-10.8(f)] in the absence of discrete valence subbands.


There are certain important similarities and differences among the 3D, 2D, 1D-nanotube electron gases, doped graphenes and carbon nanotubes in terms of the low-frequency elementary excitations and dynamic plasma waves. For any small transferred momenta, the excitation frequency is linearly proportional to the ${\bf q}$-magnitude, regardless of energy dispersions and dimensionalities \cite{Phys.Rev.B34;979}. The special structures in the bare response functions directly reflect the non-vertical joint van Hove singularities due to the initial or final Fermi-momentum/ band-edge states \cite{IOPConcisePhysics.53;15493 , Phys.Rev.B75;155430}. Generally speaking, the real-part dielectric function exhibits a huge negative value under a specific excitation frequency and then approaches to zero in the absence of electron-hole pair excitations. A prominent peak, which mainly arise from the coherent oscillations of free conduction electrons/valence holes in the $n$-/$p$-type metals, is clearly shown in each small-$q$ energy loss spectrum. At ${q\rightarrow\,0}$, the square of its frequency [${\omega_0^2}$] possesses a linear relation with the free carrier density. Apparently, the main features of charge screenings are greatly diversified in the above-mentioned five systems. For example, 1D electron gas and carbon nanotubes show more vacant regions, without any electron-hole excitations, mainly owing to only two discrete Fermi momenta in each lower-energy conduction subbands. ${Im[\epsilon\,]}$s of the above-mentioned five systems, respectively, display the cusp, asymmetric peak [strongly modified cusp], plateau [sharp peak under a narrow case], asymmetric peak in the square-root divergence, and almost symmetric peak. The specific zero point of ${Re[\epsilon\,]}$, which comes to exist at the higher frequency in the absence of ${Im[\epsilon\,]}$, leads to a prominent plasmon peak in each energy loss spectrum. The coherent oscillation frequency of all free carriers obviously reveals the momentum dependences: a constant value plus $C$$q^2$ , $\sqrt q$ , ${q\sqrt {|ln(qr)|}}$ , $\sqrt q$ and ${q\sqrt {|ln(qr)|}}$  \cite{Phys.Rev.B26;4421}. Such collective excitations experience the Landau damping beyond the critical momenta, being, respectively, induced by the intraband, intraband, intraband, inter-$\pi$-band, and inter-$\pi$-band electron-hole excitations. As for the intensity of plasmon peak, a simple relation among 3D, 2D and 1D-nanotube gases is absent, while it is lower for carbon-related honeycomb lattices. This further illustrates that the single-particle excitations of both intraband and interband excitations also affect the main features of plasmon modes through the real-part dielectric function. When the free carrier densities are sufficient high in 1D-nanotube electron gas and carbon nanotubes, there also exist few acoustic plasmon modes, respectively, due to the collective excitations of the separate conduction electrons. The latter even present the discrete inter-$\pi$-band plasmon modes at  higher frequencies [optical plasmons; \cite{PhysicaELowDimens.Syst.Nanostruct.54;267}], since the transverse quantum confinement can induce several $\pi$-valence subbands. The up-to-date measurements of energy loss spectra have successfully examined ${\omega_p(q)}$s in doped 3D semiconductors/metals [\cite{J.Phys.C:SolidStatePhys.16;6221}/ \cite{Ultramicroscopy96;469}],  2D quantum wells \cite{NatureCommunications9;2254}, 1D quantum wires \cite{PhysicaELowDimens.Syst.Nanostruct.79;20} and graphene systems \cite{Phys.Rev.B34;979} except for 1D carbon nanotubes.


\vskip 0.6 truecm
\par\noindent
{\bf 5. Concluding }
\vskip 0.3 truecm

The 3D-, 2D-, 1D-electron gases, monolayer graphene, and single-walled carbon nanotubes with free conduction electrons, are quite different from one another in terms of Coulomb excitations and propagating   plasma waves. The third and fifth systems have energy subbands due to the transverse quantum confinement, and the fourth and fifth ones possess thhe $\pi$-electron valence bands. Very interesting, the bare and screened response functions are greatly diversified by the various dimensionalities, lattice symmetries, and carrier densities. At zero temperature, the lower-frequency electronic excitations present the special structures in the dielectric functions, being mainly determined by the joint van Hove singularities associated the initial or final Fermi-momentum/band-edge states. The strong response of dynamic charge screenings clearly illustrates the discontinuous electron distribution below and above the Fermi surface. The main features of $\epsilon$s, the form, number, frequency and intensity of special structures, are very sensitive to the changes in the transferred momentum/angular momentum and  Fermi level. Both 3D and 2D parabolic conduction bands only present the non-vertical intraband excitation; therefore, the single-particle excitations could survive within the upper and lower boundaries of ${E_q\pm\,k_Fq/m}$. However, the 1D-nantube electron gas has some occupied conduction subbands and thus creates the multi-excitation channels [more electron-hole pair excitations regions]. Furthermore,   there are certain vacant regions, being frequently accompanied with the undamped and weak plasmon modes\cite{Phys.Rev.B47;6617 , Phys.Rev.B48;5567 , Chin.J.Phys.32;879}. As to the $n$-type graphene-related materials, the intra- and inter-$\pi$-band excitation channels come to exist simultaneously, thus, leading to the undamped and damped collective excitations, respectively, at the small and sufficiently large momentum transfers. Under the long wavelength limit, the coherent charge oscillations of all conduction electrons in the above-mentioned five condensed-matter systems, respectively, exhibit the rich momentum-dependent frequencies: a constant value, ${\sqrt q}$ , ${q|ln(qr)|^{1/2}}$ , ${\sqrt q}$ \cite{Phys.Rev.B26;4421} and ${q|ln(qr)|^{1/2}}$/a finite value [the ${L=0}$/ ${L=1}$ modes; \cite{Phys.Rev.B47;6617 , Phys.Rev.B48;5567 , Chin.J.Phys.32;879}]. The critical mechanisms cover the dimension-dominated bare Coulomb potentials, energy dispersions of conduction bands, and free carrier densities, being directly reflected in the real- and imaginary-part dielectric functions. Furthermore, such plasmon modes start to experience the Landau dampings after crossing the highest intraband/ the lowest interband electron-hole boundaries for electron gases/graphene-related honeycomb lattices.  In generally, two characteristics, the prominent peaks of the energy loss spectrum and the specific zero points of the real-part dielectric function, are available in identifying the distinct plasmon-mode frequencies. However, the second method is only suitable for condensed-matter systems with simple band structures, such as, 3D, 2D and 1D electron gases. Such technique becomes rather difficult for the emergent materials in the presence of rich intrinsic interactions/complicated band structures, mainly owing to the absence of ${Re[\epsilon\,]=0}$. For example, the $\pi$- and $\sigma$-plasmon modes, which could survive in graphene-related honeycomb lattices\cite{J.Appl.Phys.106;113711}, might not present the vanishing real-part dielectric function at frequencies beyond 5 and and 20 eV , respectively. There also exist the multi-acoustic plasmon modes due to certain occupied conduction subands in 1D electron gas and carbon nanotube\cite{Phys.Rev.B47;6617 , Phys.Rev.B48;5567 , Chin.J.Phys.32;879 , Phys.Rev.B50;17744}. Specifically, the latter exhibits the discrete inter-$pi$-band plasmons (optical modes, mainly owing to the occupied valence subbands).



\newpage
\renewcommand{\baselinestretch}{0.2}

\newpage

\begin{figure}[tbp]
\par
\begin{center}
\leavevmode
\includegraphics[width=1.0\linewidth]{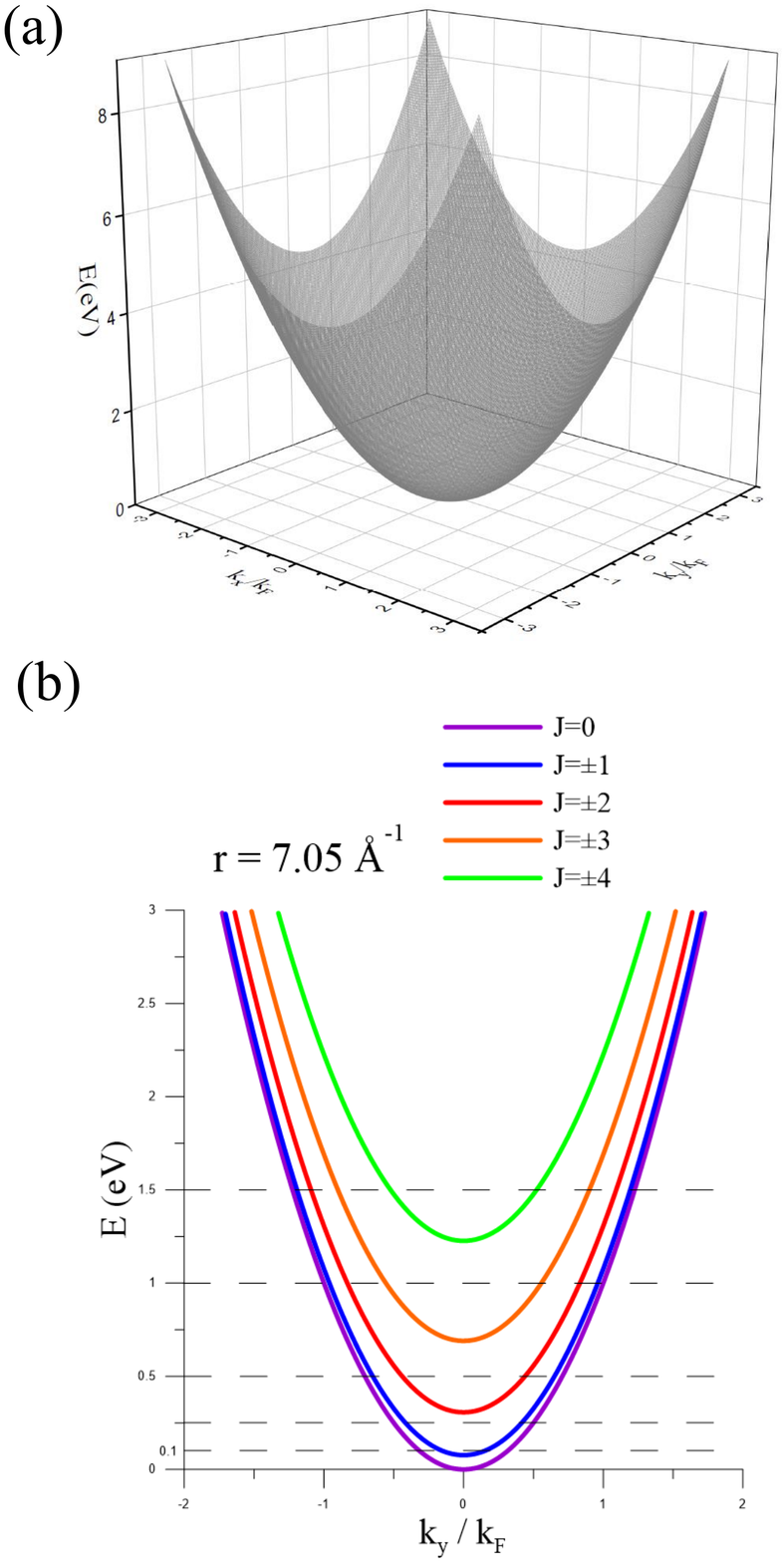}
\end{center}
\par
\textbf{fig10.1} Band structures of (a) 3D/2D and (b) 1D-nanotube electron gases at the Fermi level of ${E_F0.5}$ eV.
\end{figure}

\newpage

\begin{figure}[tbp]
\par
\begin{center}
\leavevmode
\includegraphics[width=1.0\linewidth]{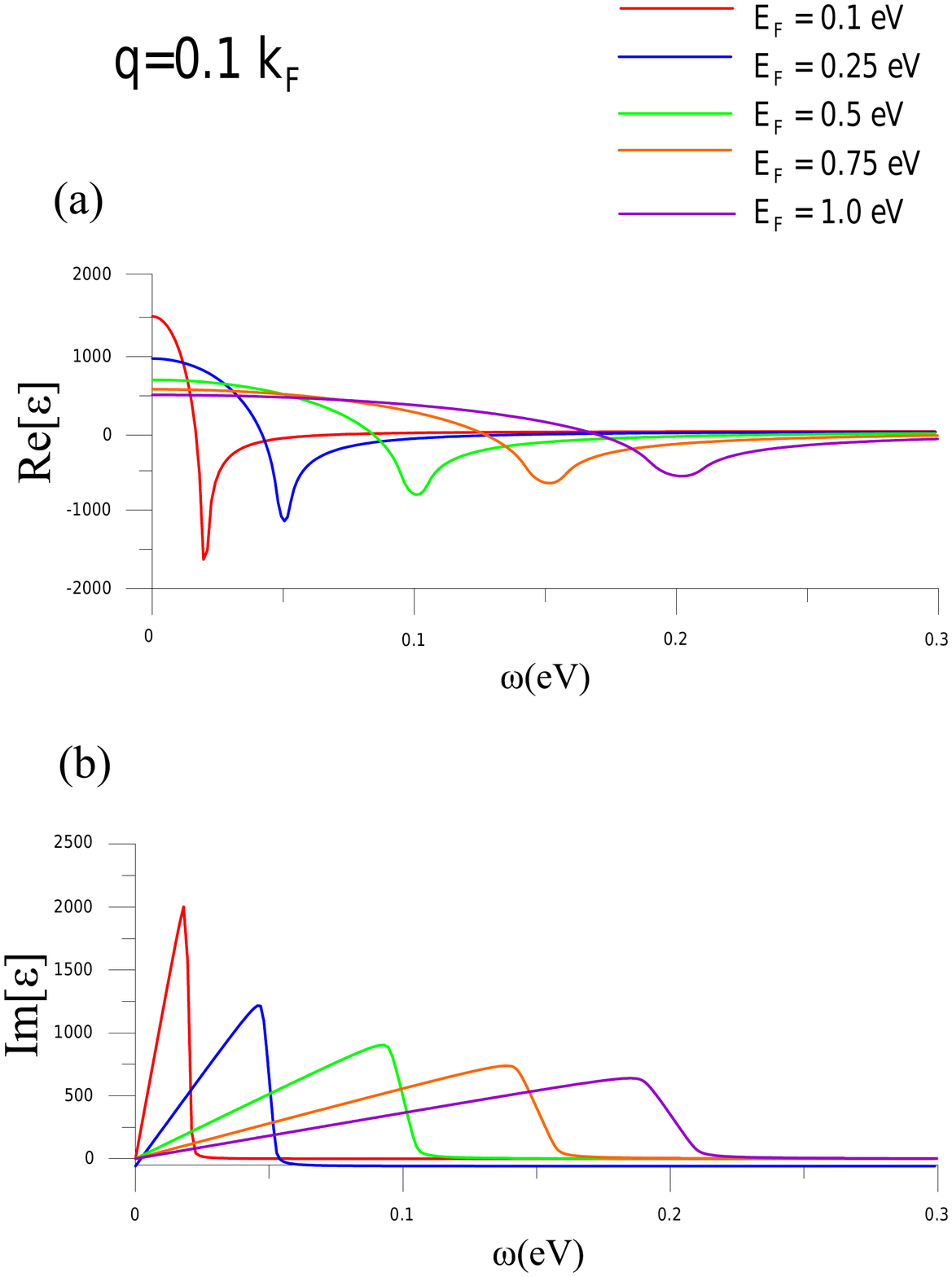}
\end{center}
\par
\textbf{fig10.2} The real- and imaginary-part dielectric functions at a specific ${q=0.1 k_F}$ and distinct Fermi levels/at a fixed ${E_F=0.5}$ eV and different momenta under the various cases: 3D electron gas (a)/(c) $\&$ (b)/(d) $\&$, 2D electron gas (e)/(g) $\&$ (f)/(h), and 1D-nanotube electron gas, with the decoupled ${L=0}$  (i)/(k) $\&$ (j)/(l) and ${L=1}$ modes (m)/(o) $\&$ (n)/(p). The insets in (c) and (d) show the specific cases of ${q=0.5}$ and 1.0 $k_F$s.
\end{figure}

\newpage

\begin{figure}[tbp]
\par
\begin{center}
\leavevmode
\includegraphics[width=1.0\linewidth]{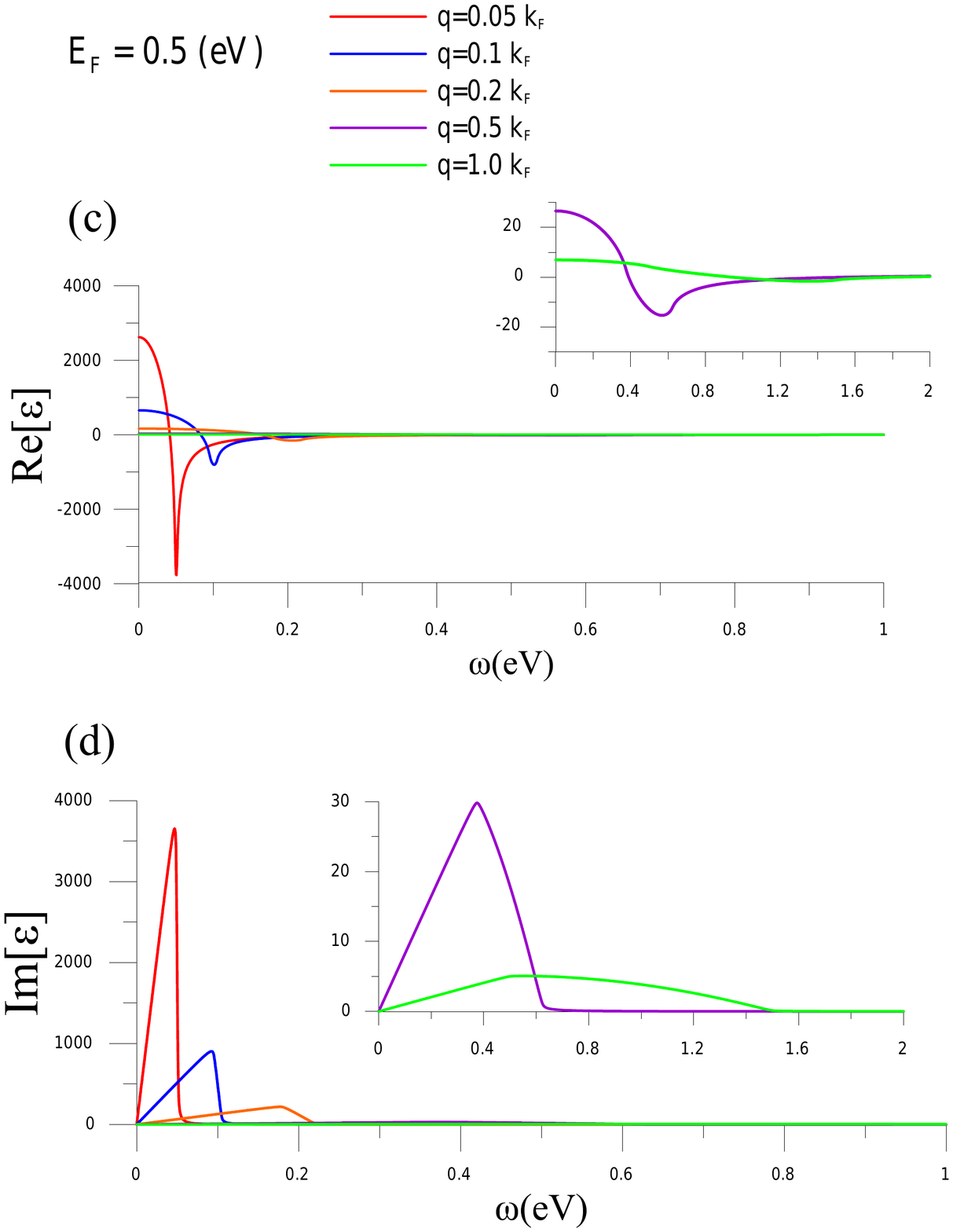}
\end{center}
\par
\textbf{fig10.2} The real- and imaginary-part dielectric functions at a specific ${q=0.1 k_F}$ and distinct Fermi levels/at a fixed ${E_F=0.5}$ eV and different momenta under the various cases: 3D electron gas (a)/(c) $\&$ (b)/(d) $\&$, 2D electron gas (e)/(g) $\&$ (f)/(h), and 1D-nanotube electron gas, with the decoupled ${L=0}$  (i)/(k) $\&$ (j)/(l) and ${L=1}$ modes (m)/(o) $\&$ (n)/(p). The insets in (c) and (d) show the specific cases of ${q=0.5}$ and 1.0 $k_F$s.
\end{figure}

\newpage

\begin{figure}[tbp]
\par
\begin{center}
\leavevmode
\includegraphics[width=1.0\linewidth]{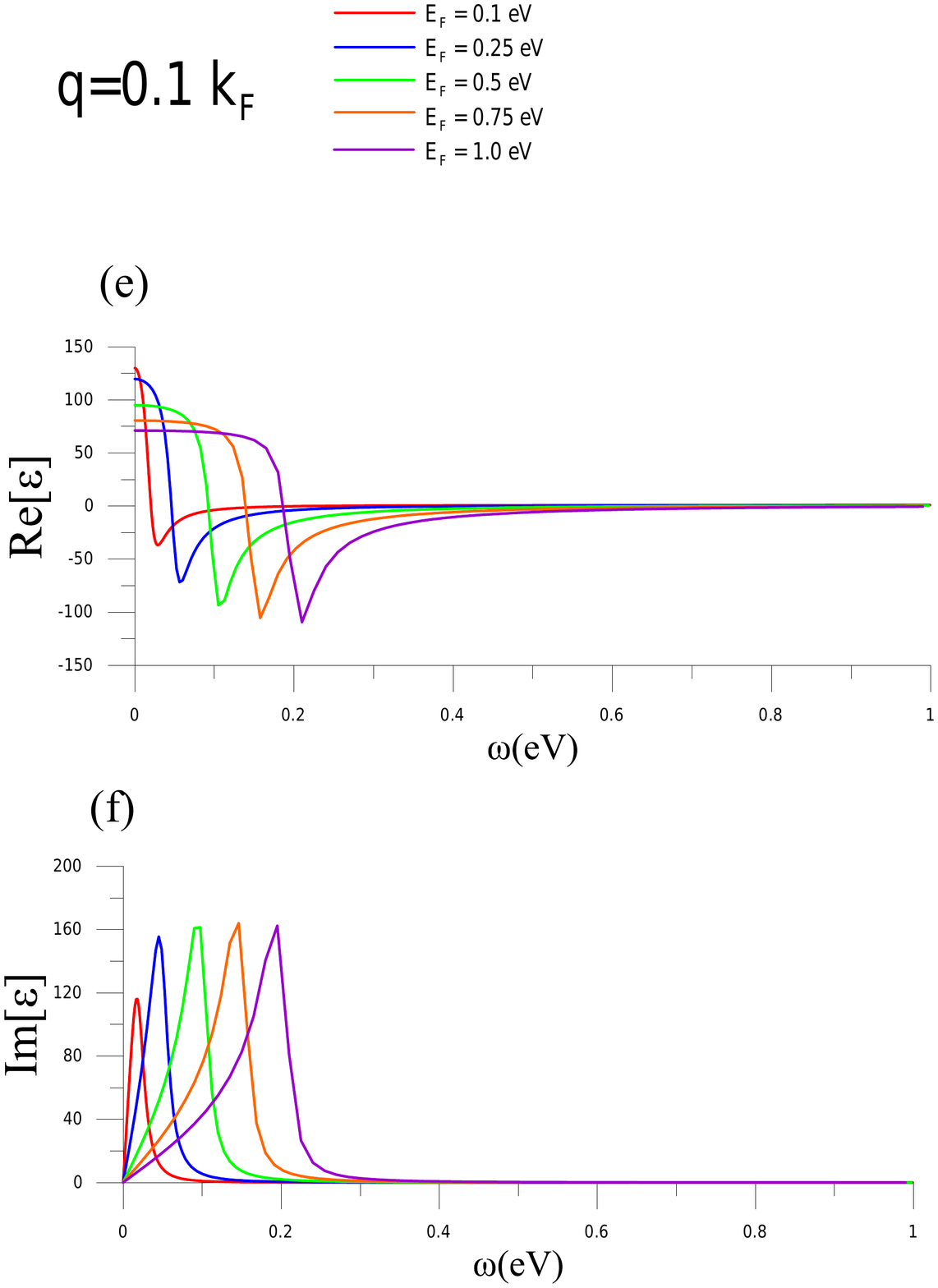}
\end{center}
\par
\textbf{fig10.2} The real- and imaginary-part dielectric functions at a specific ${q=0.1 k_F}$ and distinct Fermi levels/at a fixed ${E_F=0.5}$ eV and different momenta under the various cases: 3D electron gas (a)/(c) $\&$ (b)/(d) $\&$, 2D electron gas (e)/(g) $\&$ (f)/(h), and 1D-nanotube electron gas, with the decoupled ${L=0}$  (i)/(k) $\&$ (j)/(l) and ${L=1}$ modes (m)/(o) $\&$ (n)/(p). The insets in (c) and (d) show the specific cases of ${q=0.5}$ and 1.0 $k_F$s.
\end{figure}

\newpage

\begin{figure}[tbp]
\par
\begin{center}
\leavevmode
\includegraphics[width=1.0\linewidth]{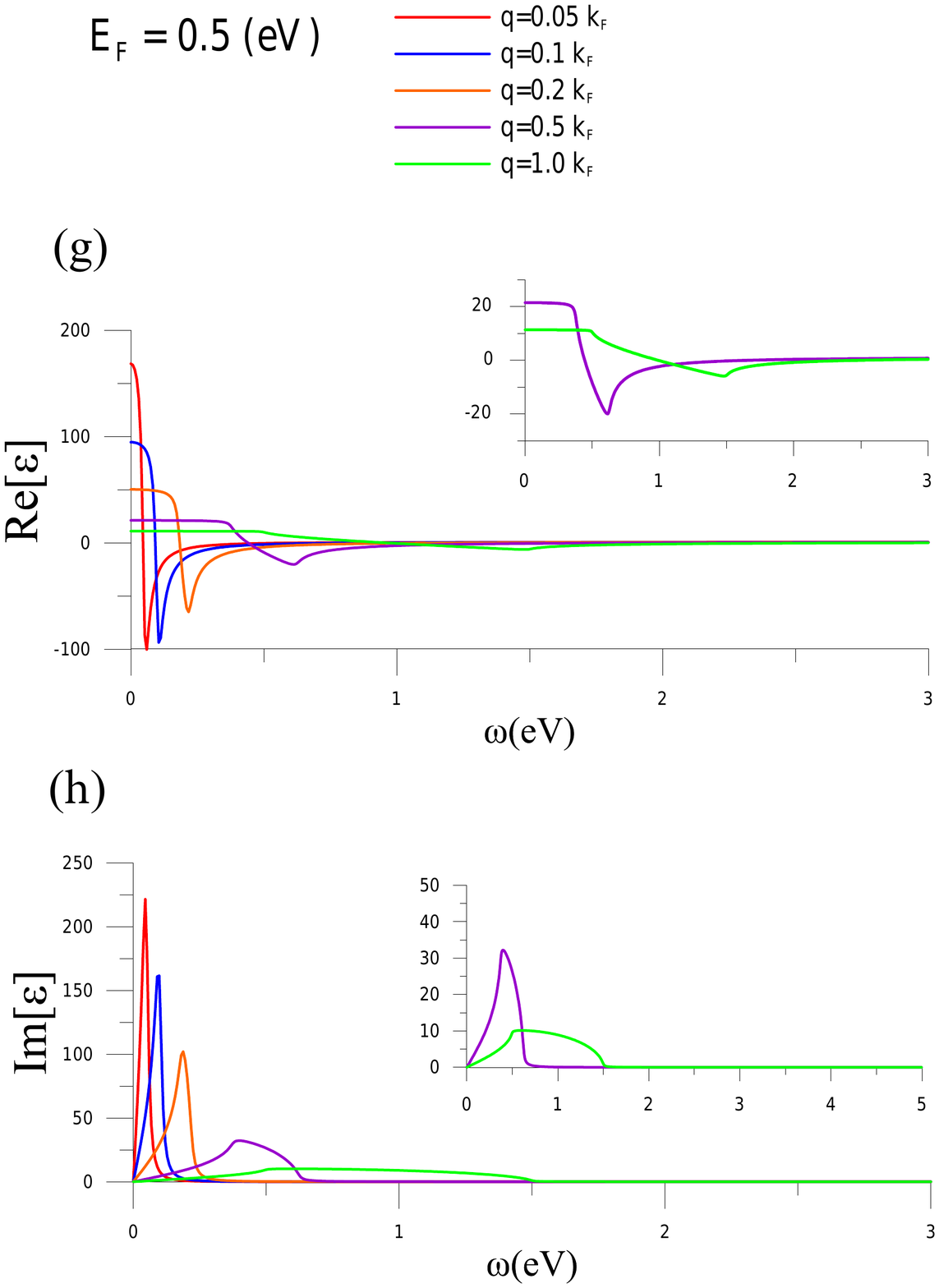}
\end{center}
\par
\textbf{fig10.2} The real- and imaginary-part dielectric functions at a specific ${q=0.1 k_F}$ and distinct Fermi levels/at a fixed ${E_F=0.5}$ eV and different momenta under the various cases: 3D electron gas (a)/(c) $\&$ (b)/(d) $\&$, 2D electron gas (e)/(g) $\&$ (f)/(h), and 1D-nanotube electron gas, with the decoupled ${L=0}$  (i)/(k) $\&$ (j)/(l) and ${L=1}$ modes (m)/(o) $\&$ (n)/(p). The insets in (c) and (d) show the specific cases of ${q=0.5}$ and 1.0 $k_F$s.
\end{figure}

\newpage

\begin{figure}[tbp]
\par
\begin{center}
\leavevmode
\includegraphics[width=1.0\linewidth]{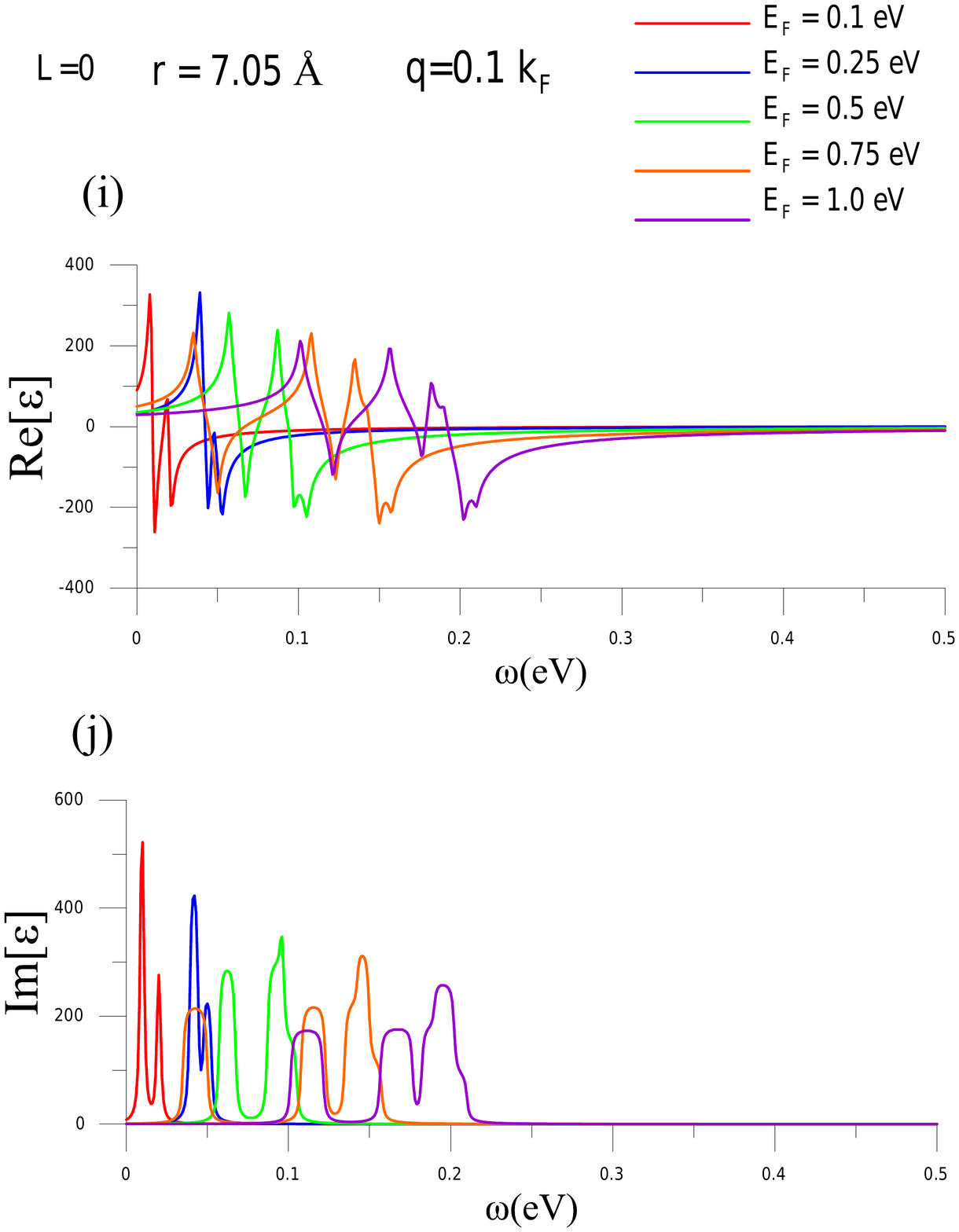}
\end{center}
\par
\textbf{fig10.2} The real- and imaginary-part dielectric functions at a specific ${q=0.1 k_F}$ and distinct Fermi levels/at a fixed ${E_F=0.5}$ eV and different momenta under the various cases: 3D electron gas (a)/(c) $\&$ (b)/(d) $\&$, 2D electron gas (e)/(g) $\&$ (f)/(h), and 1D-nanotube electron gas, with the decoupled ${L=0}$  (i)/(k) $\&$ (j)/(l) and ${L=1}$ modes (m)/(o) $\&$ (n)/(p). The insets in (c) and (d) show the specific cases of ${q=0.5}$ and 1.0 $k_F$s.
\end{figure}

\newpage

\begin{figure}[tbp]
\par
\begin{center}
\leavevmode
\includegraphics[width=1.0\linewidth]{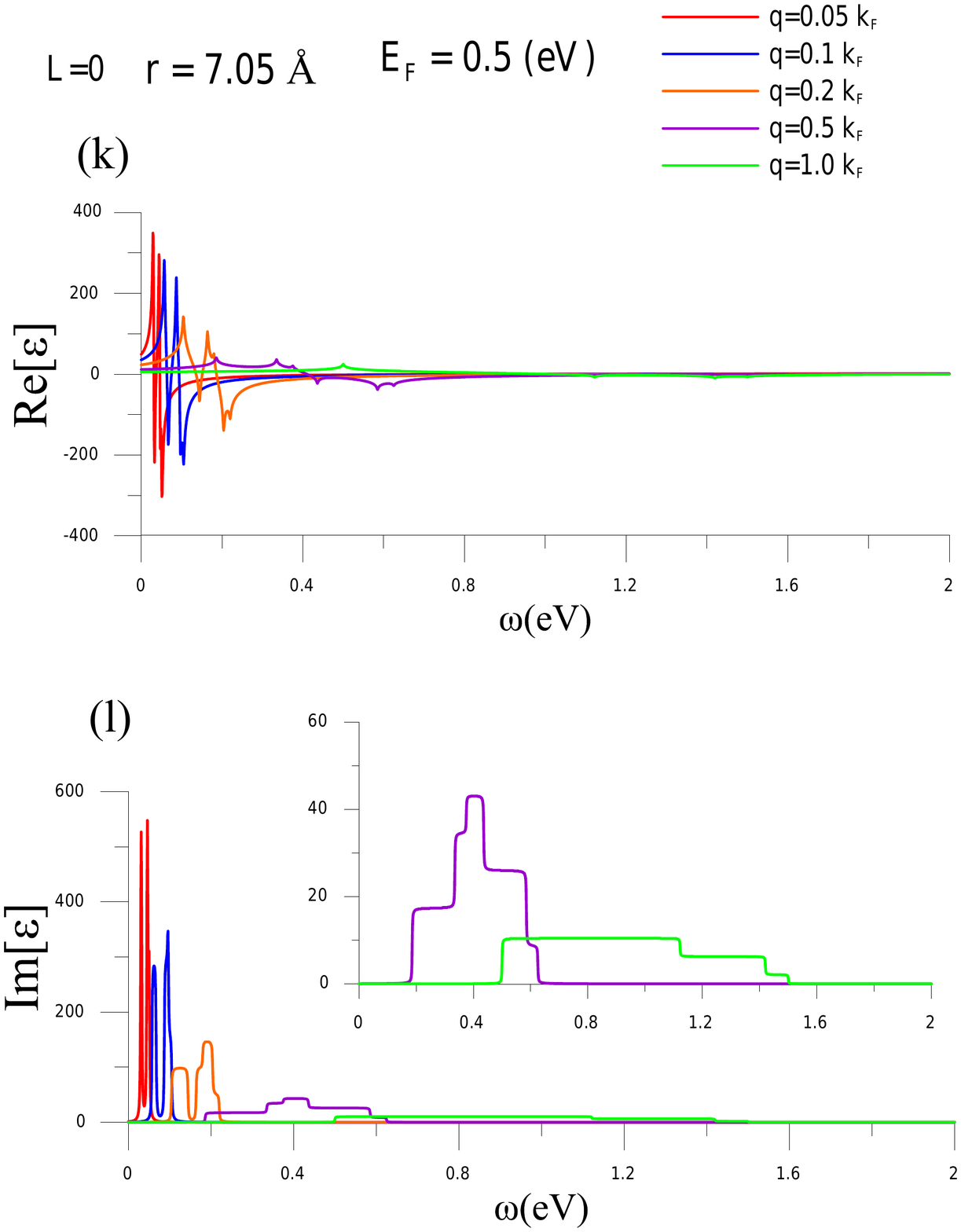}
\end{center}
\par
\textbf{fig10.2} The real- and imaginary-part dielectric functions at a specific ${q=0.1 k_F}$ and distinct Fermi levels/at a fixed ${E_F=0.5}$ eV and different momenta under the various cases: 3D electron gas (a)/(c) $\&$ (b)/(d) $\&$, 2D electron gas (e)/(g) $\&$ (f)/(h), and 1D-nanotube electron gas, with the decoupled ${L=0}$  (i)/(k) $\&$ (j)/(l) and ${L=1}$ modes (m)/(o) $\&$ (n)/(p). The insets in (c) and (d) show the specific cases of ${q=0.5}$ and 1.0 $k_F$s.
\end{figure}

\newpage

\begin{figure}[tbp]
\par
\begin{center}
\leavevmode
\includegraphics[width=1.0\linewidth]{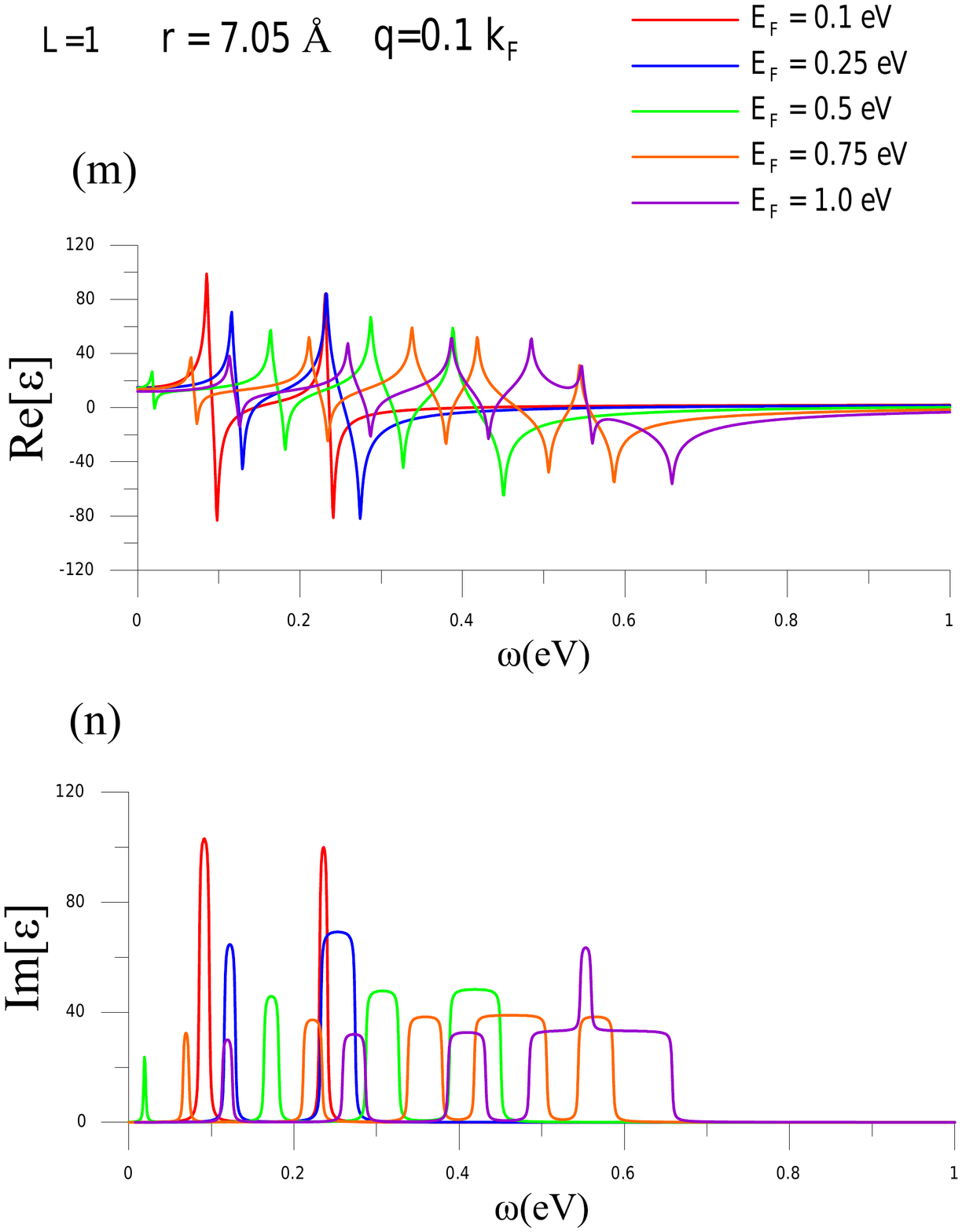}
\end{center}
\par
\textbf{fig10.2} The real- and imaginary-part dielectric functions at a specific ${q=0.1 k_F}$ and distinct Fermi levels/at a fixed ${E_F=0.5}$ eV and different momenta under the various cases: 3D electron gas (a)/(c) $\&$ (b)/(d) $\&$, 2D electron gas (e)/(g) $\&$ (f)/(h), and 1D-nanotube electron gas, with the decoupled ${L=0}$  (i)/(k) $\&$ (j)/(l) and ${L=1}$ modes (m)/(o) $\&$ (n)/(p). The insets in (c) and (d) show the specific cases of ${q=0.5}$ and 1.0 $k_F$s.
\end{figure}

\newpage

\begin{figure}[tbp]
\par
\begin{center}
\leavevmode
\includegraphics[width=1.0\linewidth]{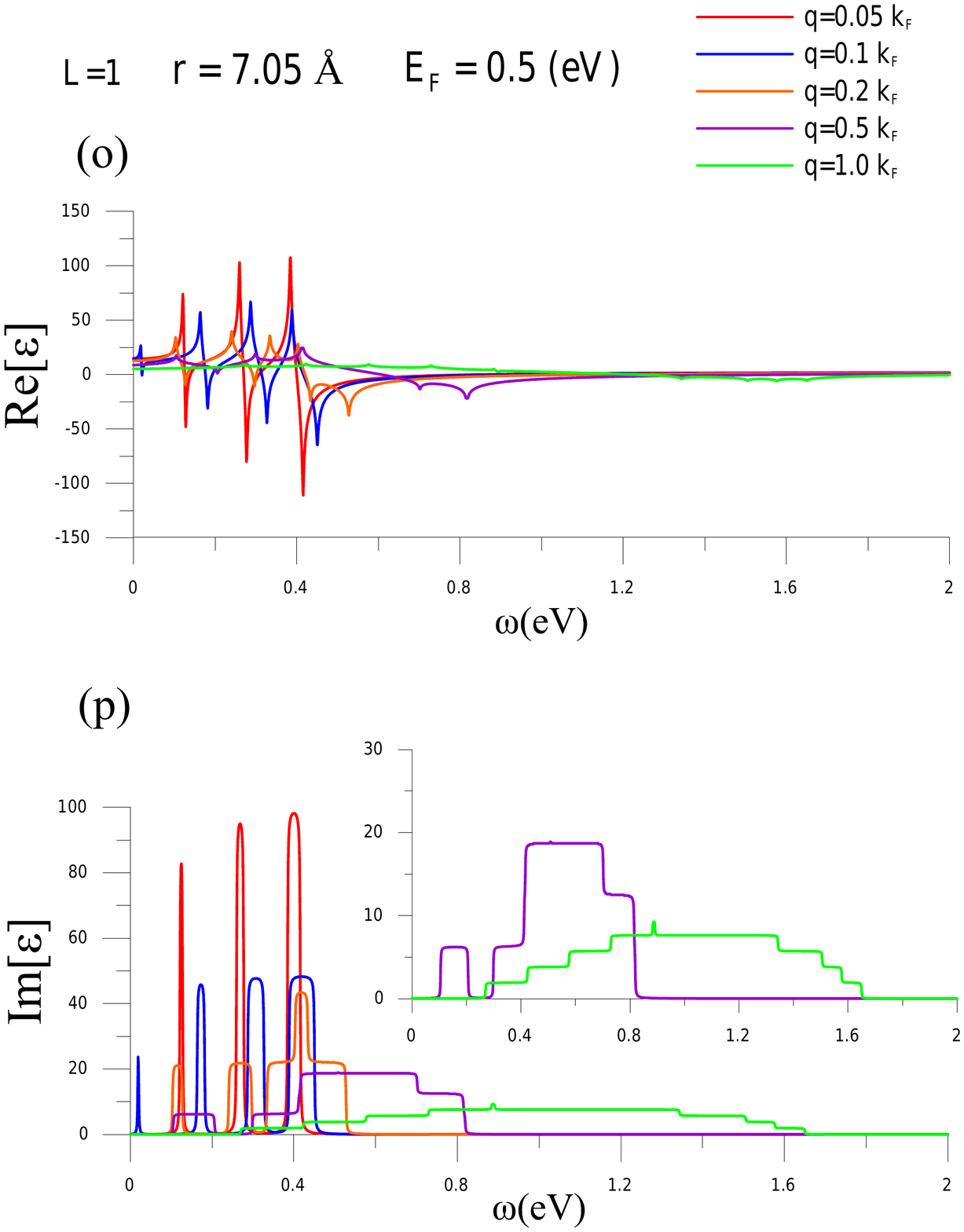}
\end{center}
\par
\textbf{fig10.2} The real- and imaginary-part dielectric functions at a specific ${q=0.1 k_F}$ and distinct Fermi levels/at a fixed ${E_F=0.5}$ eV and different momenta under the various cases: 3D electron gas (a)/(c) $\&$ (b)/(d) $\&$, 2D electron gas (e)/(g) $\&$ (f)/(h), and 1D-nanotube electron gas, with the decoupled ${L=0}$  (i)/(k) $\&$ (j)/(l) and ${L=1}$ modes (m)/(o) $\&$ (n)/(p). The insets in (c) and (d) show the specific cases of ${q=0.5}$ and 1.0 $k_F$s.
\end{figure}

\newpage

\begin{figure}[tbp]
\par
\begin{center}
\leavevmode
\includegraphics[width=1.0\linewidth]{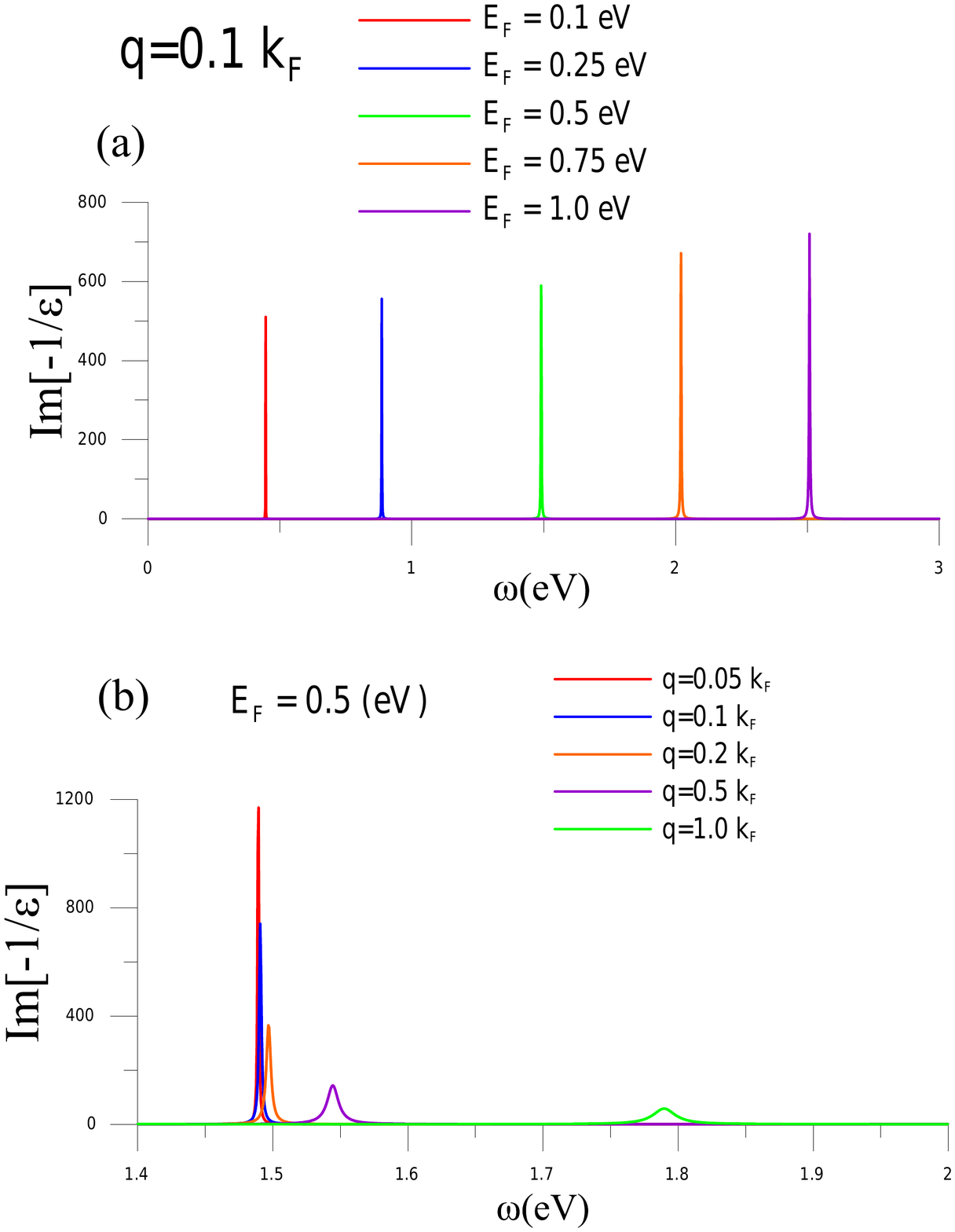}
\end{center}
\par
\textbf{fig10.3} The energy loss spectra at ${q=0.1}$ $k_F$ $\&$ distinct Fermi levesl/under ${E_F=0.5}$ eV $\&$ various momentum transfers  for (a)/(b) 3D, (c)/(d) 2D, and 1D-nanotube electron gases, respectively, corresponding to the (e)/(f) ${L=0}$ and (g)/(h) ${L=1}$ normal channels.
\end{figure}

\newpage

\begin{figure}[tbp]
\par
\begin{center}
\leavevmode
\includegraphics[width=1.0\linewidth]{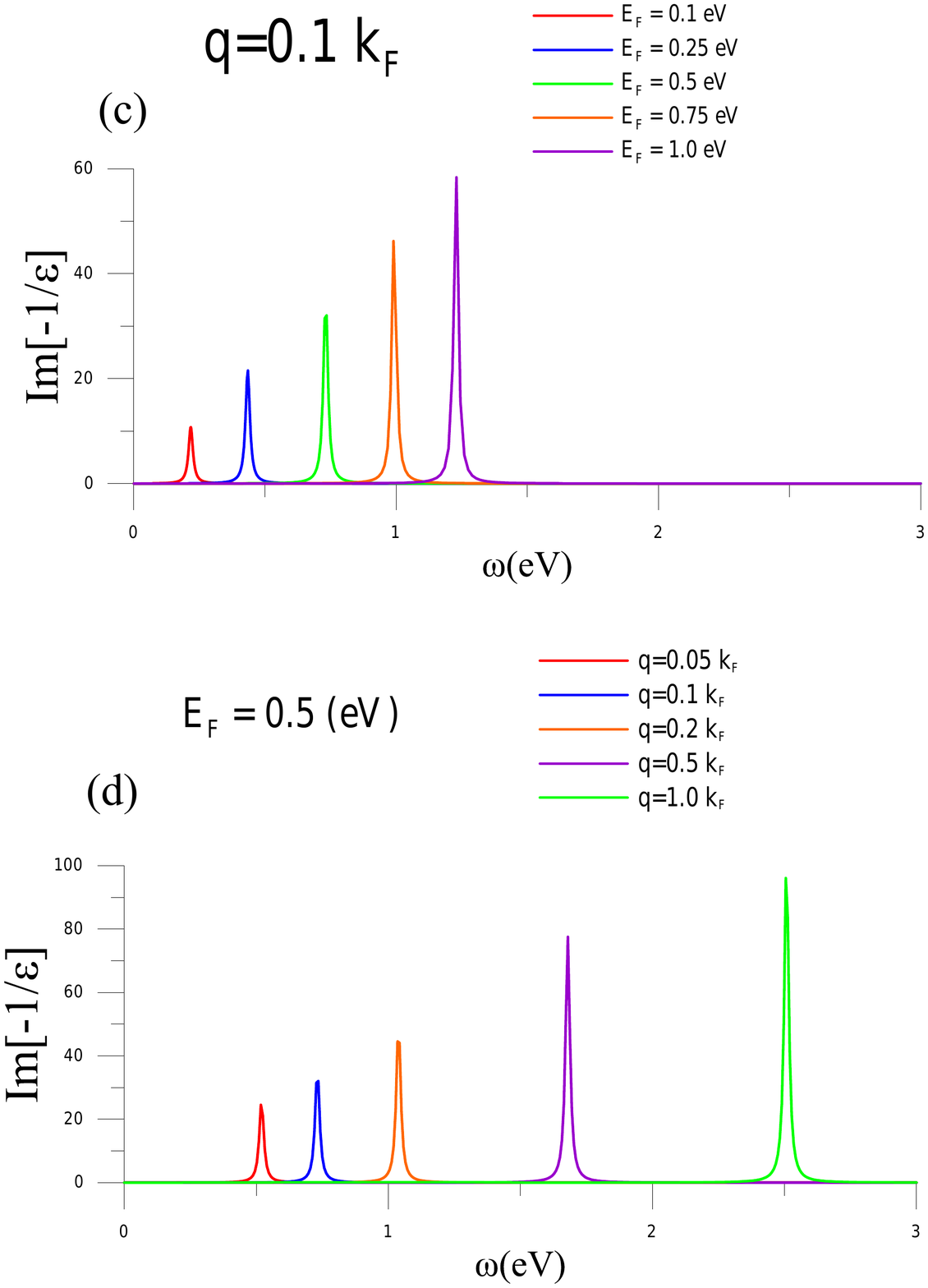}
\end{center}
\par
\textbf{fig10.3} The energy loss spectra at ${q=0.1}$ $k_F$ $\&$ distinct Fermi levesl/under ${E_F=0.5}$ eV $\&$ various momentum transfers  for (a)/(b) 3D, (c)/(d) 2D, and 1D-nanotube electron gases, respectively, corresponding to the (e)/(f) ${L=0}$ and (g)/(h) ${L=1}$ normal channels.
\end{figure}

\newpage

\begin{figure}[tbp]
\par
\begin{center}
\leavevmode
\includegraphics[width=1.0\linewidth]{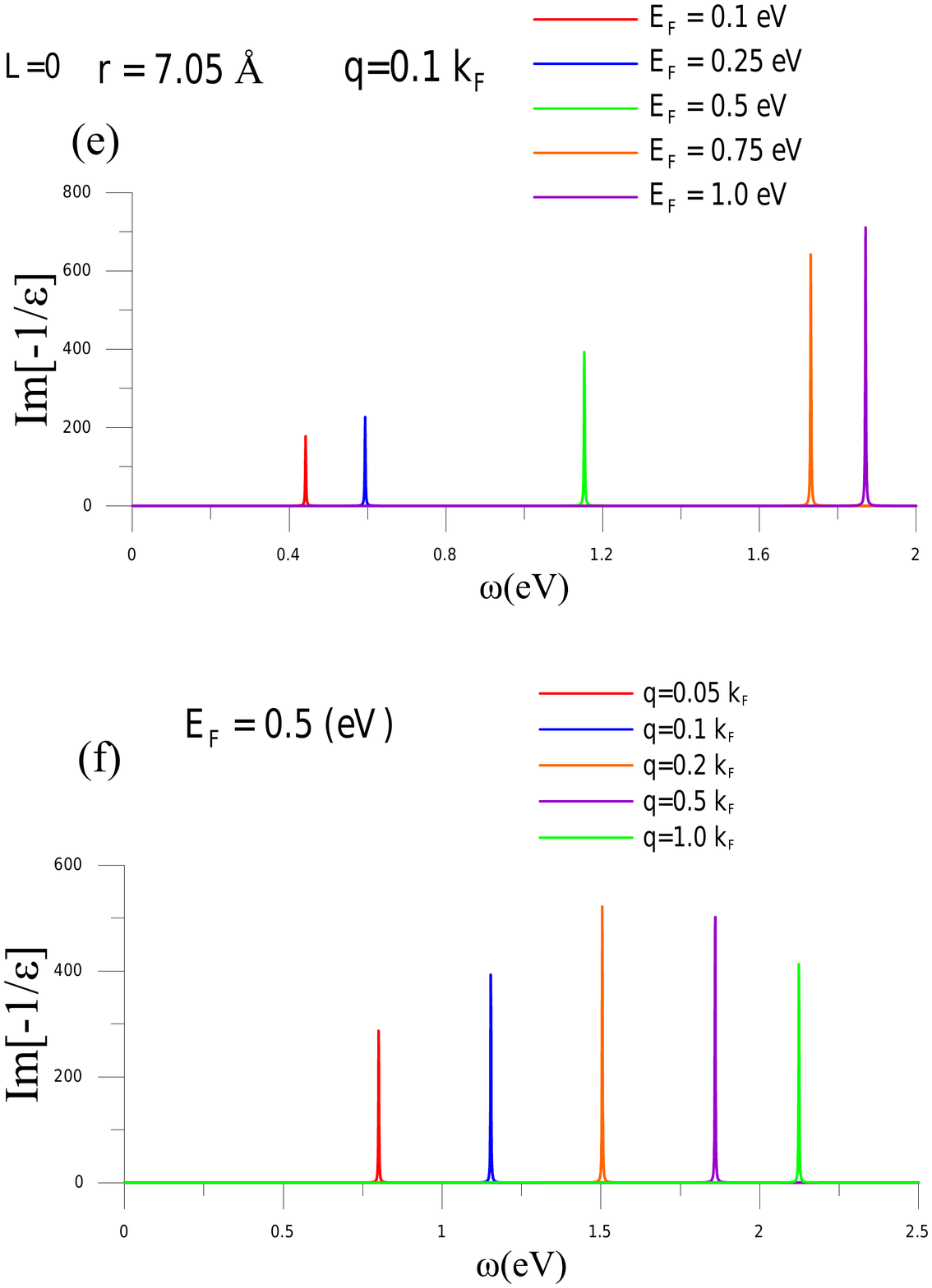}
\end{center}
\par
\textbf{fig10.3} The energy loss spectra at ${q=0.1}$ $k_F$ $\&$ distinct Fermi levesl/under ${E_F=0.5}$ eV $\&$ various momentum transfers  for (a)/(b) 3D, (c)/(d) 2D, and 1D-nanotube electron gases, respectively, corresponding to the (e)/(f) ${L=0}$ and (g)/(h) ${L=1}$ normal channels.
\end{figure}

\newpage

\begin{figure}[tbp]
\par
\begin{center}
\leavevmode
\includegraphics[width=1.0\linewidth]{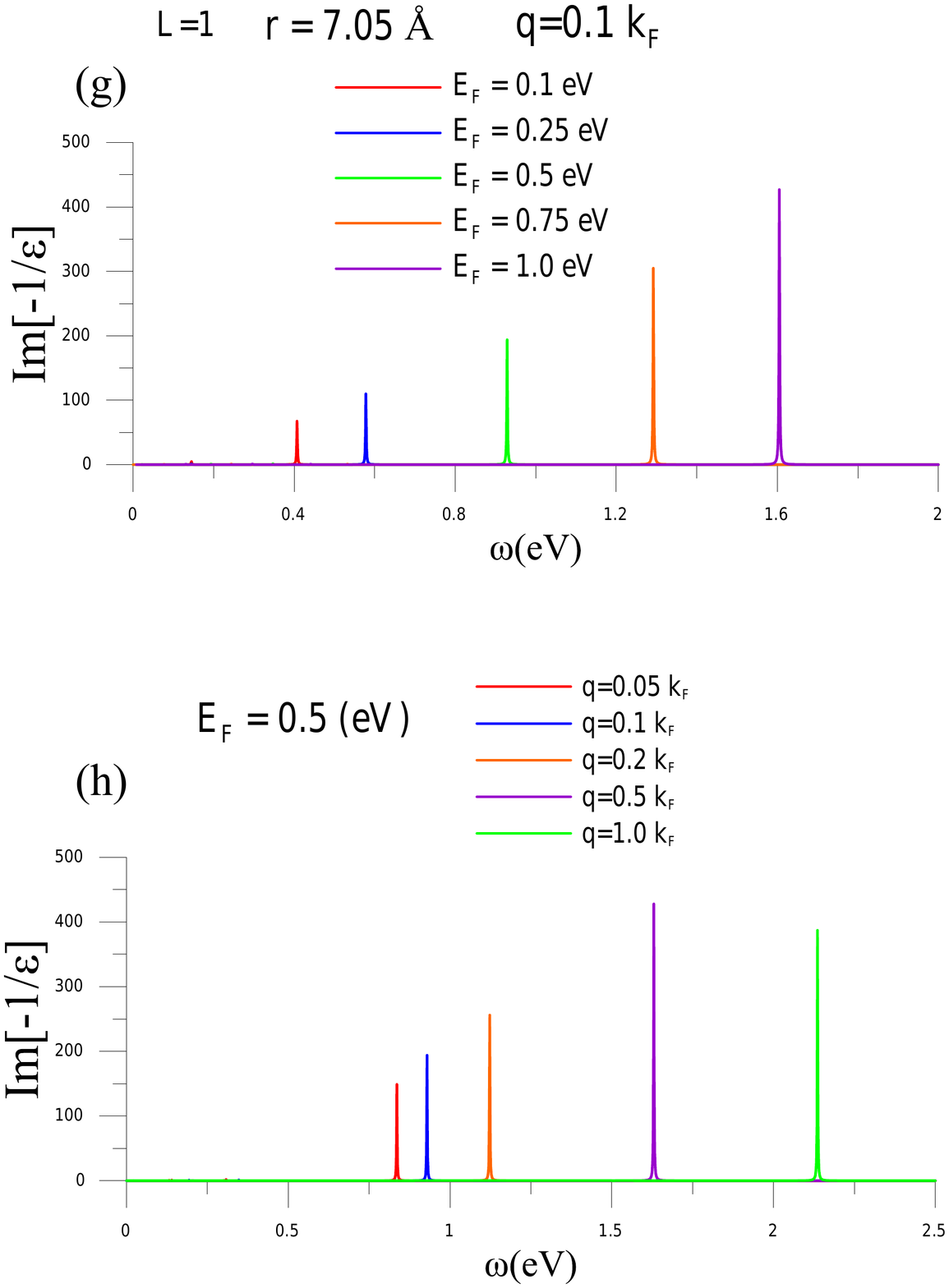}
\end{center}
\par
\textbf{fig10.3} The energy loss spectra at ${q=0.1}$ $k_F$ $\&$ distinct Fermi levesl/under ${E_F=0.5}$ eV $\&$ various momentum transfers  for (a)/(b) 3D, (c)/(d) 2D, and 1D-nanotube electron gases, respectively, corresponding to the (e)/(f) ${L=0}$ and (g)/(h) ${L=1}$ normal channels.
\end{figure}

\newpage

\begin{figure}[tbp]
\par
\begin{center}
\leavevmode
\includegraphics[width=1.0\linewidth]{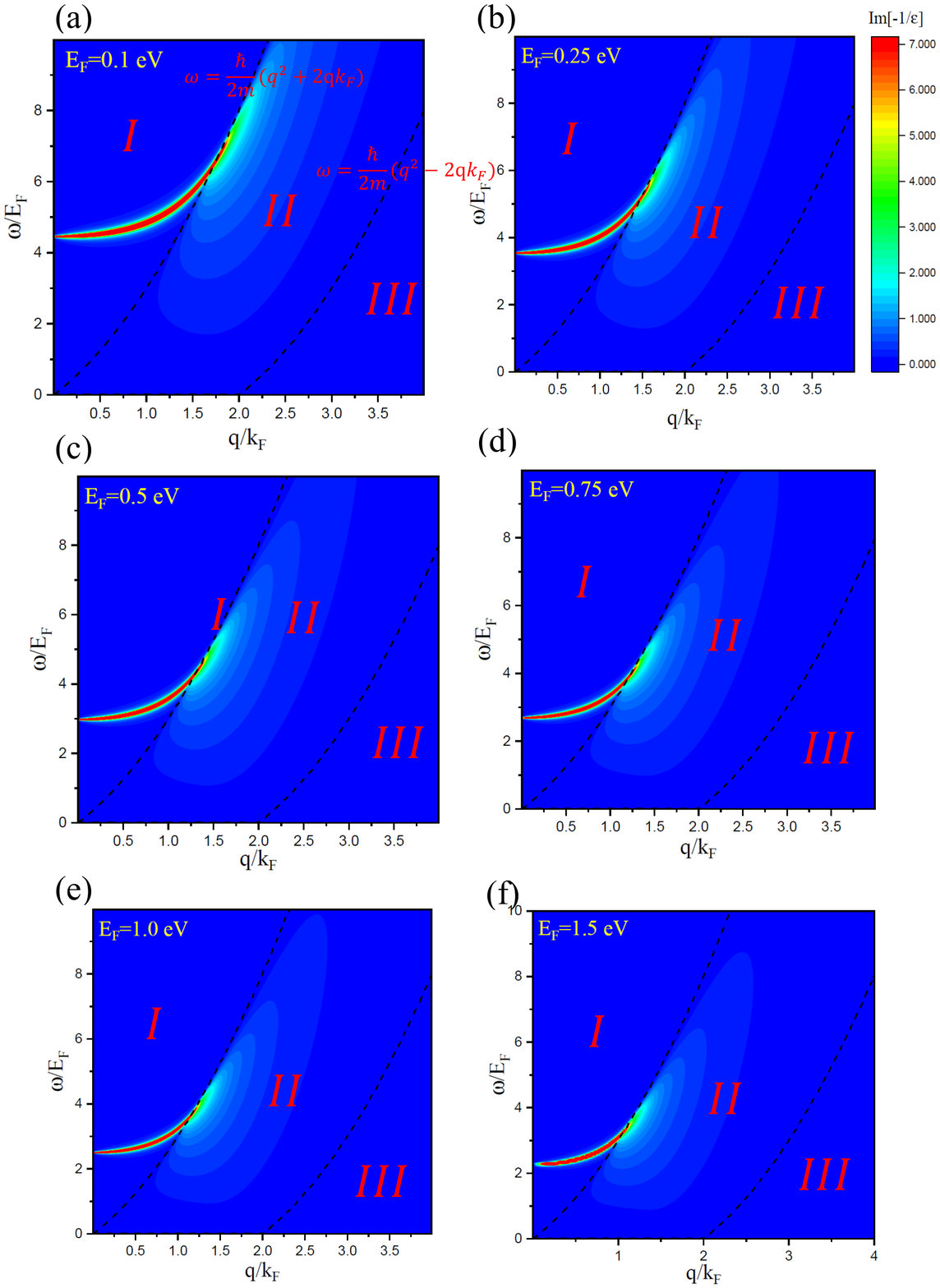}
\end{center}
\par
\textbf{fig10.4} The [momentum, frequency]-excitation phase diagrams under the different Fermi energies of           ${E_F}$s=0.10/0.25/0.5/0.75/1 and 1.5 eVs  for (a)/(b)/(c)/(d)/(e)/(f) 3D, (g)/(h)/(i)/(j)/(k)/(l) 2D and only 1D electron gas with the normal modes of (m)/(n)/(o)/(p)/(q)/(r) ${L=0}$ and (s)/(t)/(u)/(v)/(w)/(x) ${L=1}$.
\end{figure}

\newpage

\begin{figure}[tbp]
\par
\begin{center}
\leavevmode
\includegraphics[width=1.0\linewidth]{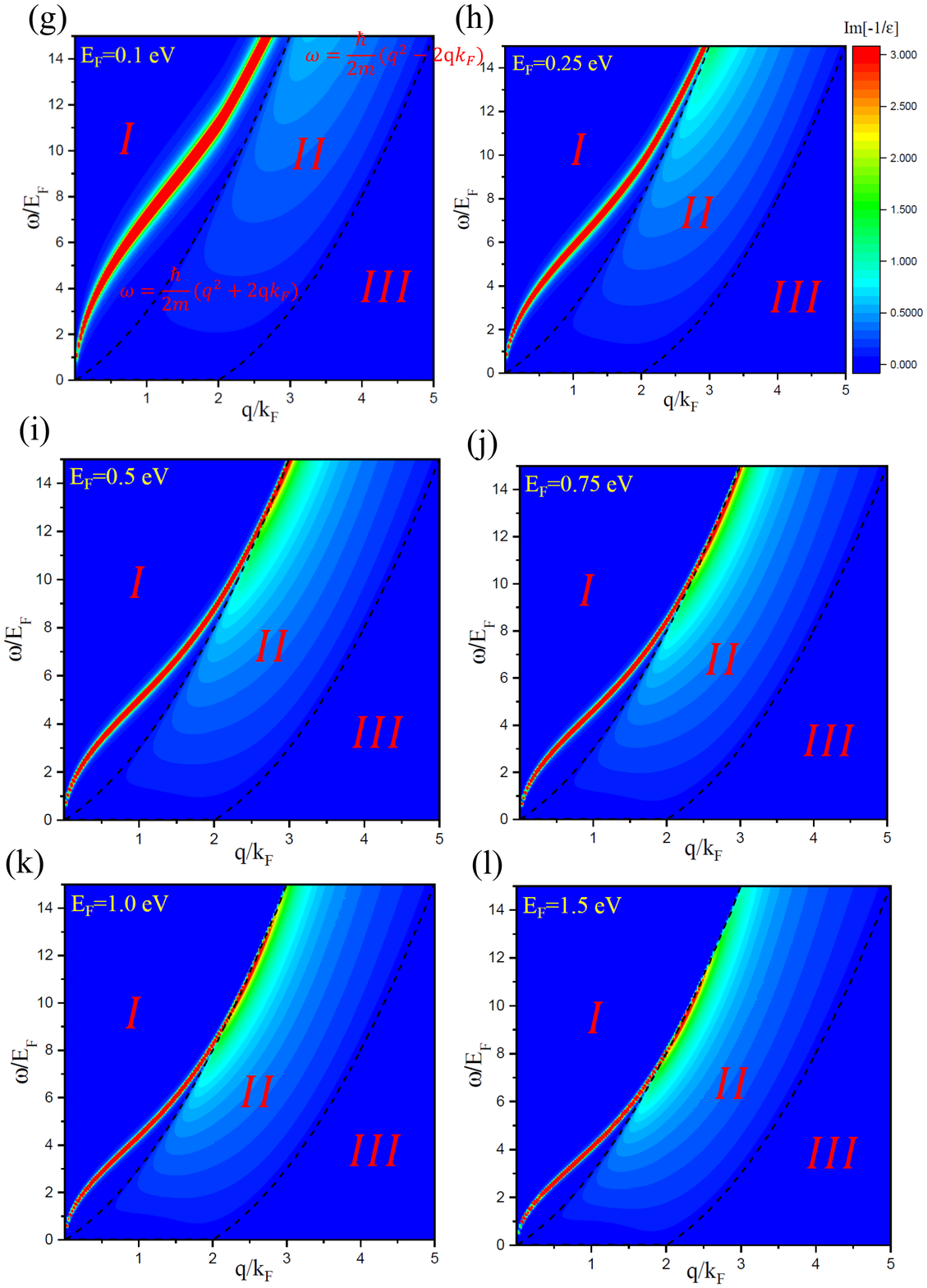}
\end{center}
\par
\textbf{fig10.4} The [momentum, frequency]-excitation phase diagrams under the different Fermi energies of           ${E_F}$s=0.10/0.25/0.5/0.75/1 and 1.5 eVs  for (a)/(b)/(c)/(d)/(e)/(f) 3D, (g)/(h)/(i)/(j)/(k)/(l) 2D and only 1D electron gas with the normal modes of (m)/(n)/(o)/(p)/(q)/(r) ${L=0}$ and (s)/(t)/(u)/(v)/(w)/(x) ${L=1}$.
\end{figure}

\newpage

\begin{figure}[tbp]
\par
\begin{center}
\leavevmode
\includegraphics[width=1.0\linewidth]{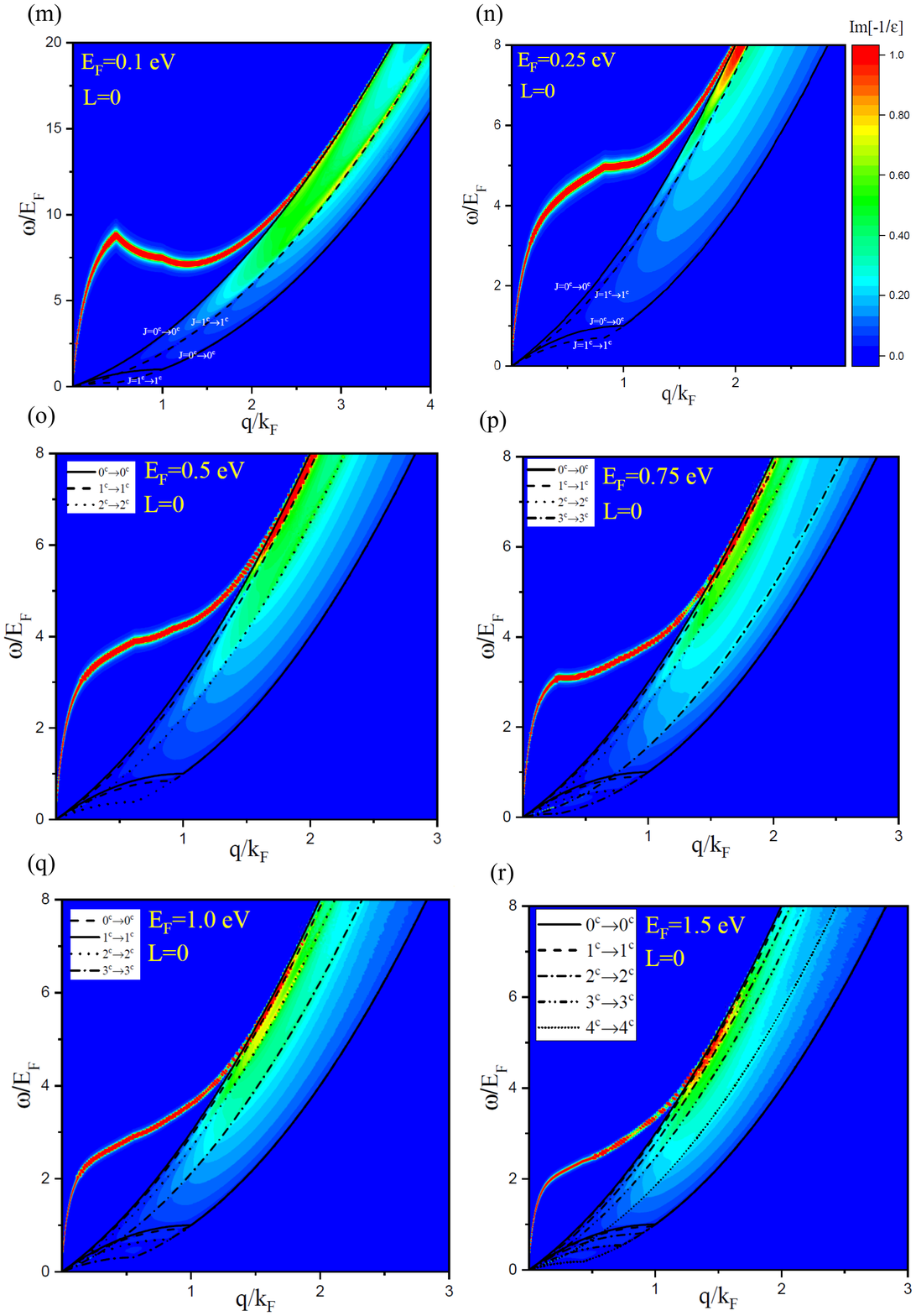}
\end{center}
\par
\textbf{fig10.4} The [momentum, frequency]-excitation phase diagrams under the different Fermi energies of           ${E_F}$s=0.10/0.25/0.5/0.75/1 and 1.5 eVs  for (a)/(b)/(c)/(d)/(e)/(f) 3D, (g)/(h)/(i)/(j)/(k)/(l) 2D and only 1D electron gas with the normal modes of (m)/(n)/(o)/(p)/(q)/(r) ${L=0}$ and (s)/(t)/(u)/(v)/(w)/(x) ${L=1}$.
\end{figure}

\newpage

\begin{figure}[tbp]
\par
\begin{center}
\leavevmode
\includegraphics[width=1.0\linewidth]{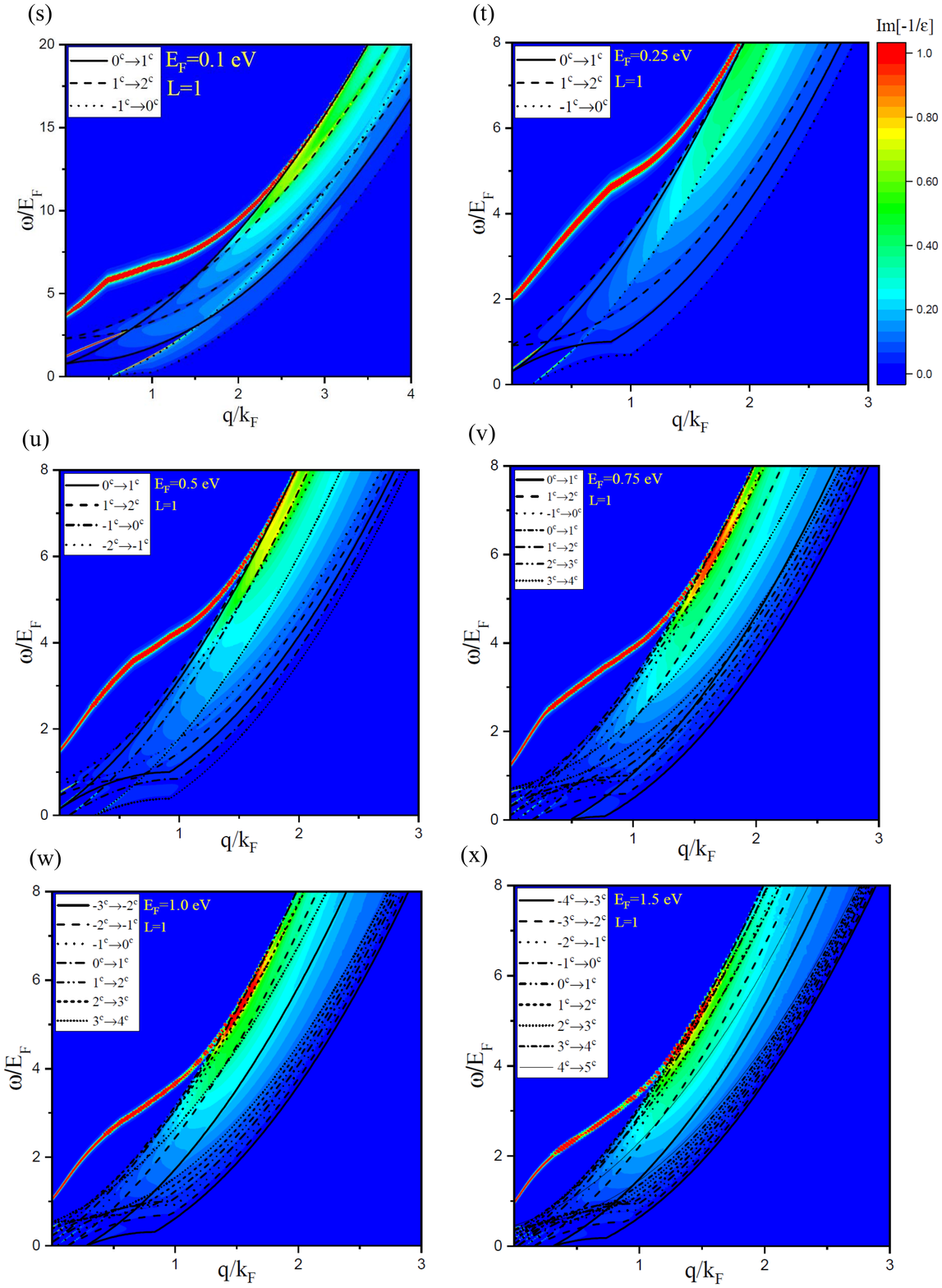}
\end{center}
\par
\textbf{fig10.4} The [momentum, frequency]-excitation phase diagrams under the different Fermi energies of           ${E_F}$s=0.10/0.25/0.5/0.75/1 and 1.5 eVs  for (a)/(b)/(c)/(d)/(e)/(f) 3D, (g)/(h)/(i)/(j)/(k)/(l) 2D and only 1D electron gas with the normal modes of (m)/(n)/(o)/(p)/(q)/(r) ${L=0}$ and (s)/(t)/(u)/(v)/(w)/(x) ${L=1}$.
\end{figure}

\newpage

\begin{figure}[tbp]
\par
\begin{center}
\leavevmode
\includegraphics[width=1.0\linewidth]{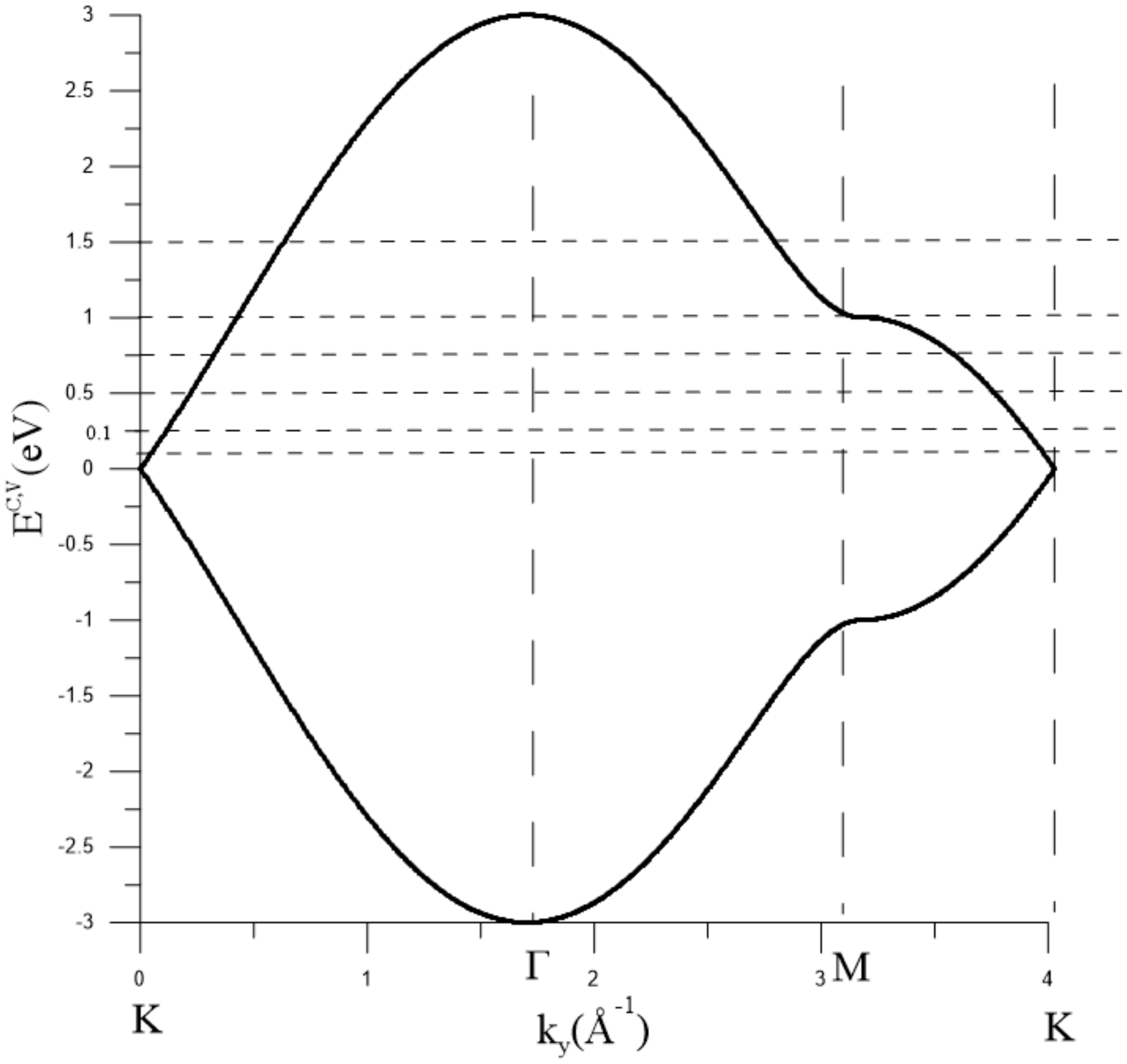}
\end{center}
\par
\textbf{fig10.5} The $\pi$-electronic structures for the doped (a) monolayer graphene and (b) zigzag (18, 0) carbon nanotube under the various Fermi levels. Two arrows indicate the intra-/inter- [${c\rightarrow\,c}$] and inter-$pi$-band [${v\rightarrow\,c}$] Coulomb excitations. The inset in (b) shows a narrow gap.
\end{figure}

\newpage

\begin{figure}[tbp]
\par
\begin{center}
\leavevmode
\includegraphics[width=1.0\linewidth]{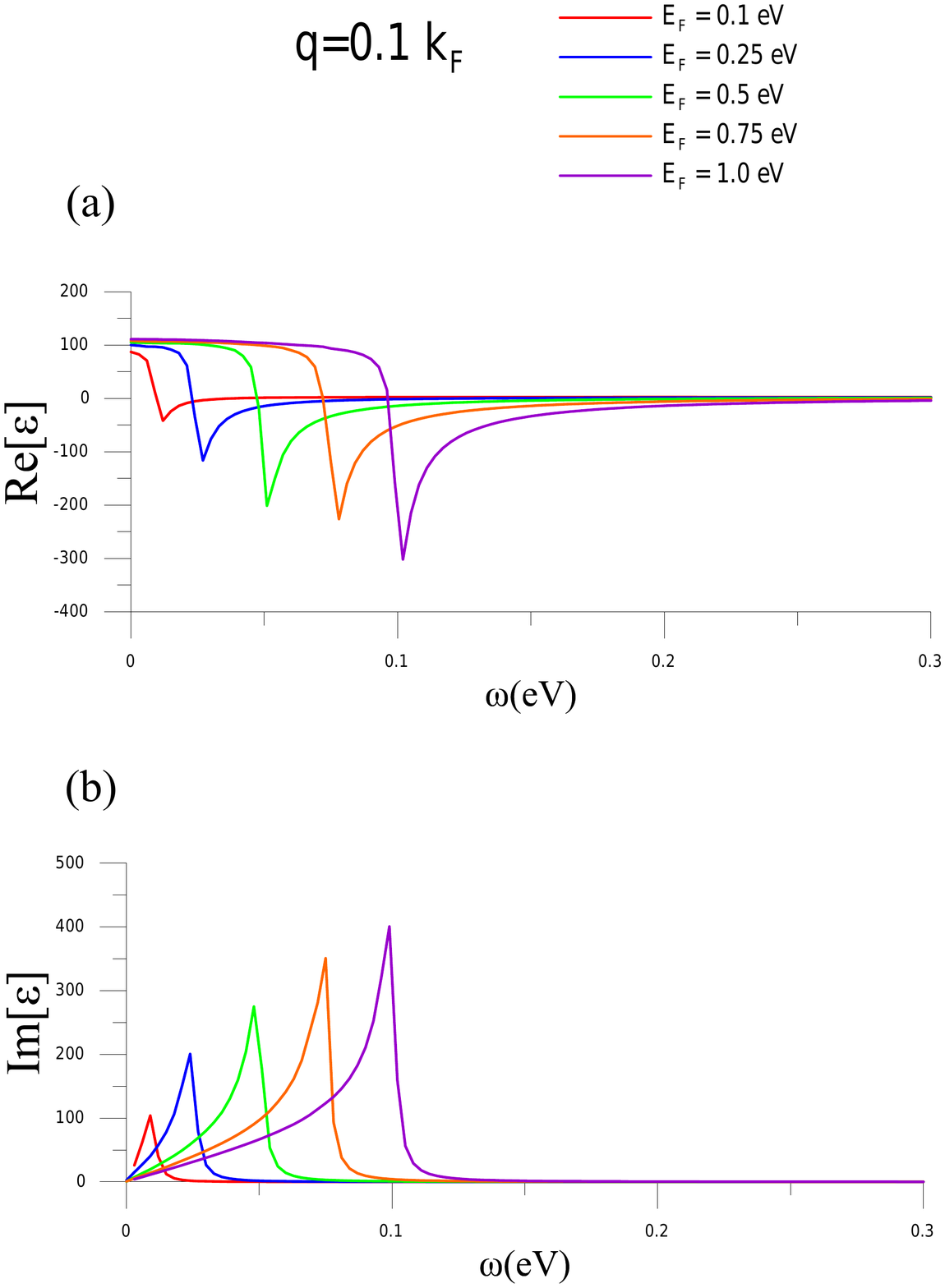}
\end{center}
\par
\textbf{fig10.6} ${Re[\epsilon\,]}$ and ${Im[\epsilon\,]}$ of a metallic graphene (a) $\&$ (b) under a specific ${q=0.1 k_F}$ and various Fermi levels/(c) $\&$ (d) at a fixed ${E_F=0.5}$ eV and distinct momenta.
\end{figure}

\newpage

\begin{figure}[tbp]
\par
\begin{center}
\leavevmode
\includegraphics[width=1.0\linewidth]{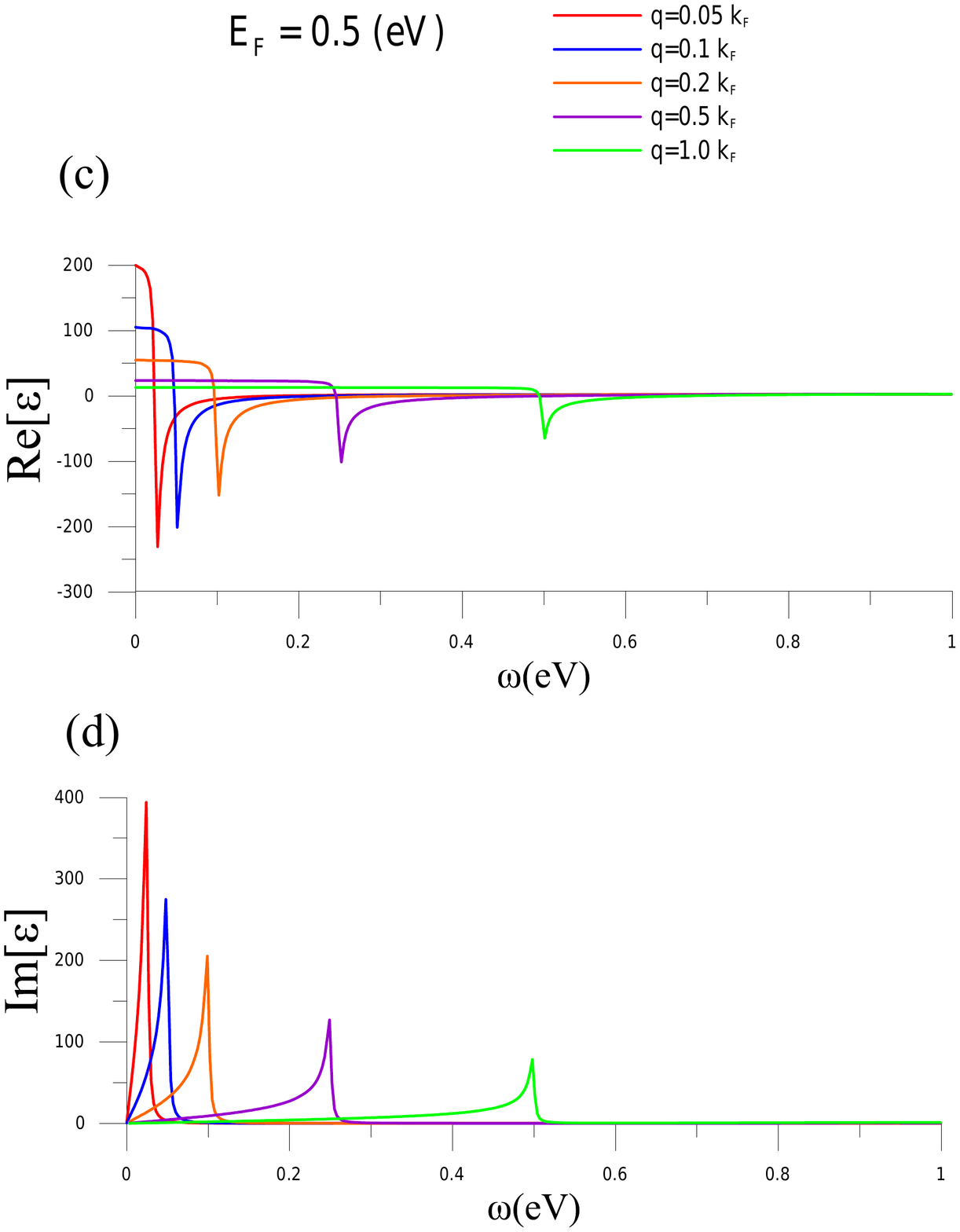}
\end{center}
\par
\textbf{fig10.6} ${Re[\epsilon\,]}$ and ${Im[\epsilon\,]}$ of a metallic graphene (a) $\&$ (b) under a specific ${q=0.1 k_F}$ and various Fermi levels/(c) $\&$ (d) at a fixed ${E_F=0.5}$ eV and distinct momenta.
\end{figure}

\newpage

\begin{figure}[tbp]
\par
\begin{center}
\leavevmode
\includegraphics[width=1.0\linewidth]{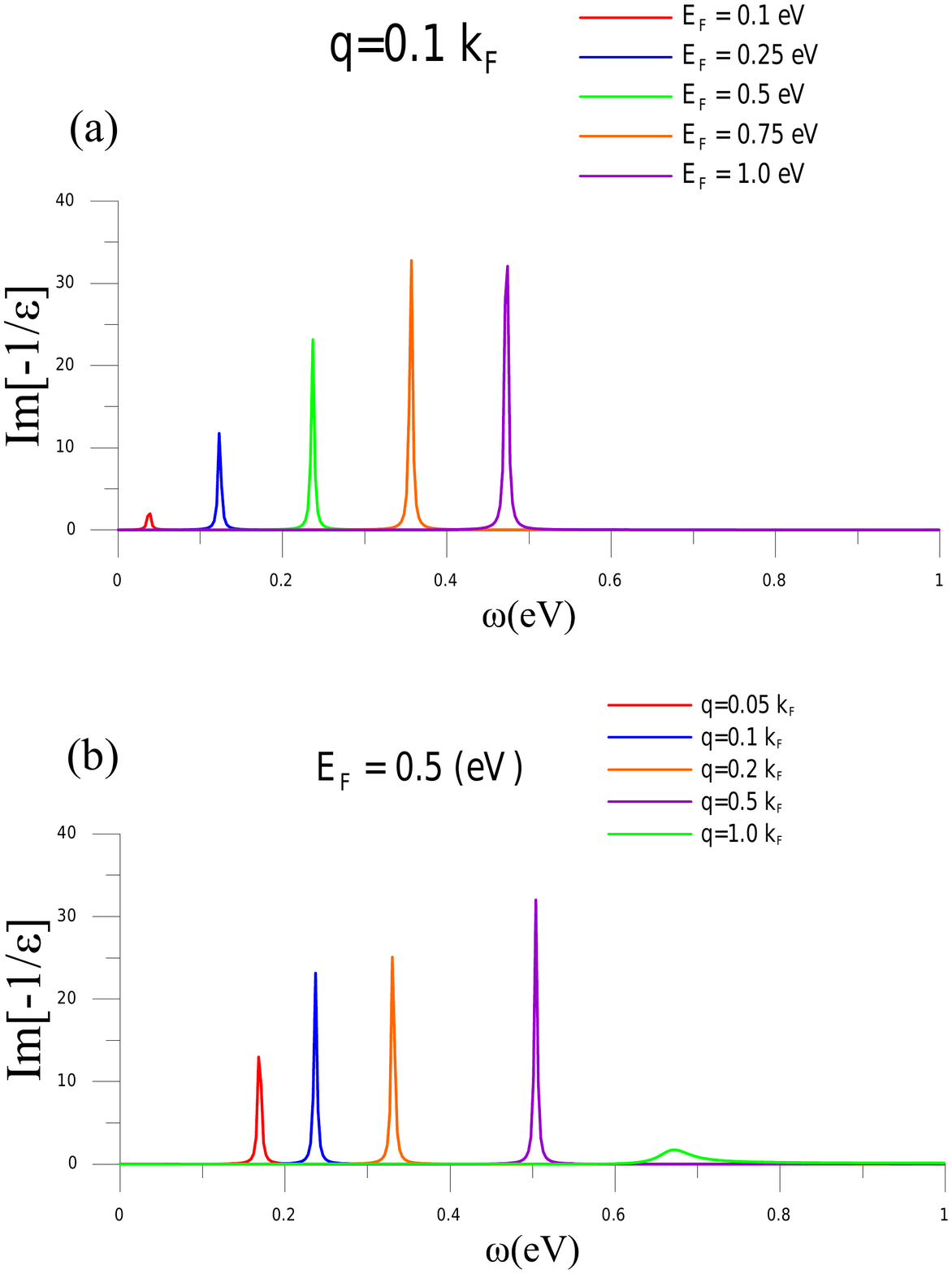}
\end{center}
\par
\textbf{fig10.7} ${Im[-1/\epsilon\,]}$s of $n$-type graphenes under (a) ${q=0.1}$ $k_F$ $\&$ various Fermi levels and (b) ${E_F=0.1}$ eV  $\&$ distinct transferred momenta.
\end{figure}

\newpage

\begin{figure}[tbp]
\par
\begin{center}
\leavevmode
\includegraphics[width=1.0\linewidth]{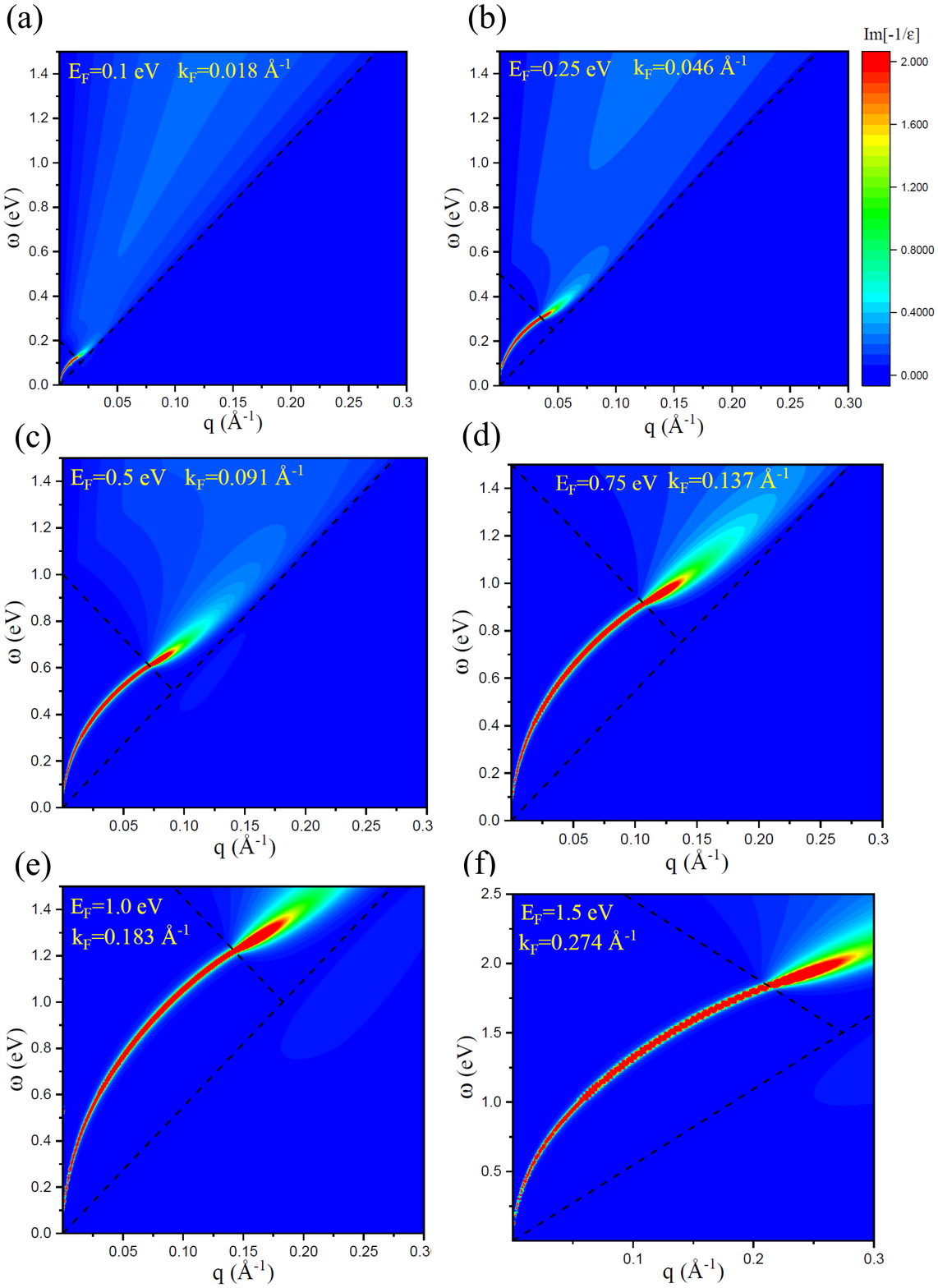}
\end{center}
\par
\textbf{fig10.8} The [${q, \omega}$]-excitation phase diagrams of metallic graphenes with various doping levels: ${E_F}$s=0.10, 0.25, 0.5, 0.75, 1.0 and 1.5 eVs, respectively, in (a), (b), (c), (d), (e) and (f).
\end{figure}

\newpage

\begin{figure}[tbp]
\par
\begin{center}
\leavevmode
\includegraphics[width=1.0\linewidth]{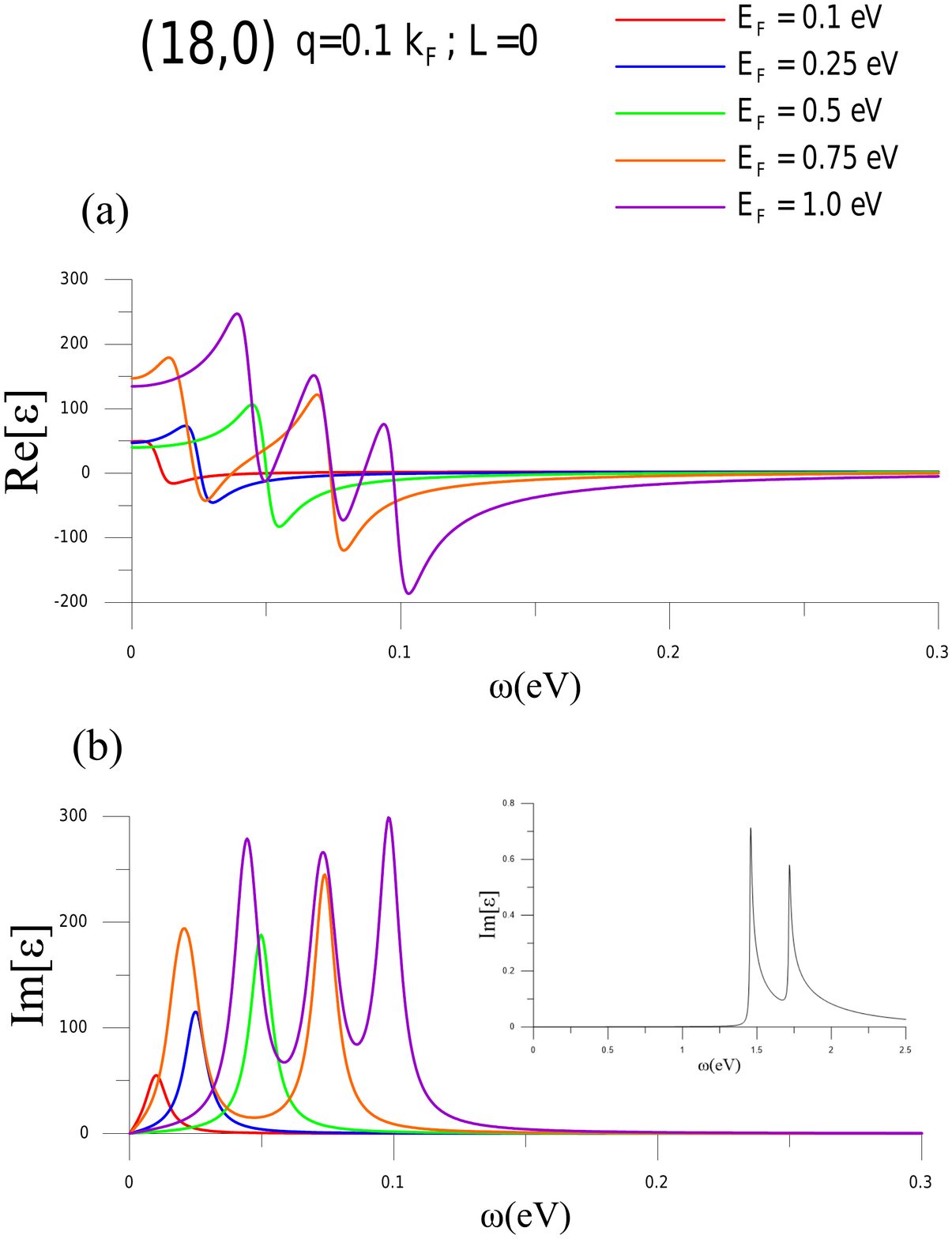}
\end{center}
\par
\textbf{fig10.9} The real-/imaginary-part dielectric function in a doped (18, 0) carbon nanotube for the ${L=0}$ mode (a)/(b) at ${q=0.1 k_F}$ and different Fermi levels, and (c)/(d) under ${E_F=0.5}$ eV and various  momentum transfers. The inset of (b) shows the inter-$pi$-band prominent response at the higher frequency beyond ${2E_F}$. The similar results of the ${L=1}$ mode are, respectively, illustrated, in (e)/(f) and (g)/(h).
\end{figure}

\newpage

\begin{figure}[tbp]
\par
\begin{center}
\leavevmode
\includegraphics[width=1.0\linewidth]{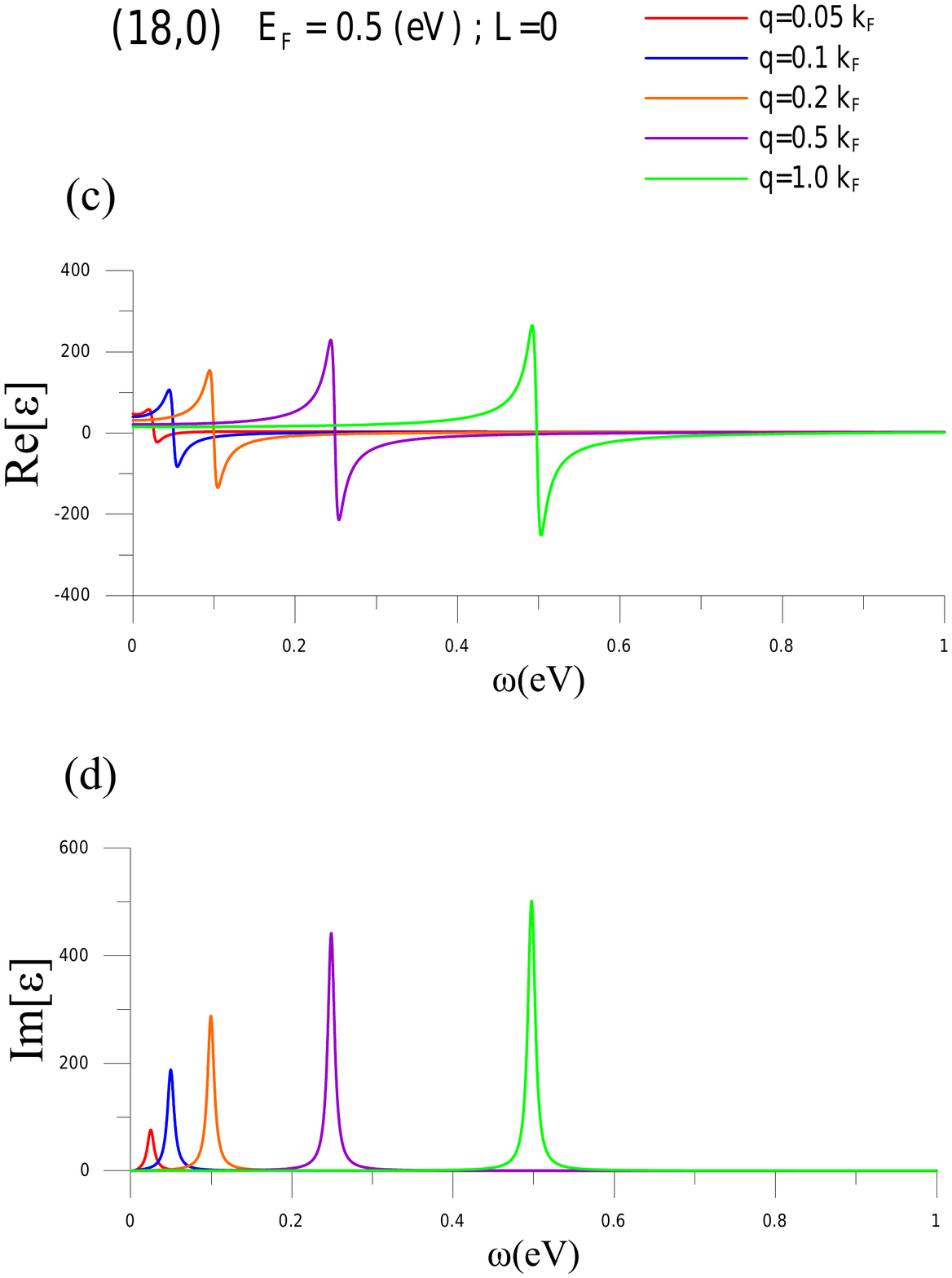}
\end{center}
\par
\textbf{fig10.9} The real-/imaginary-part dielectric function in a doped (18, 0) carbon nanotube for the ${L=0}$ mode (a)/(b) at ${q=0.1 k_F}$ and different Fermi levels, and (c)/(d) under ${E_F=0.5}$ eV and various  momentum transfers. The inset of (b) shows the inter-$pi$-band prominent response at the higher frequency beyond ${2E_F}$. The similar results of the ${L=1}$ mode are, respectively, illustrated, in (e)/(f) and (g)/(h).
\end{figure}

\newpage

\begin{figure}[tbp]
\par
\begin{center}
\leavevmode
\includegraphics[width=1.0\linewidth]{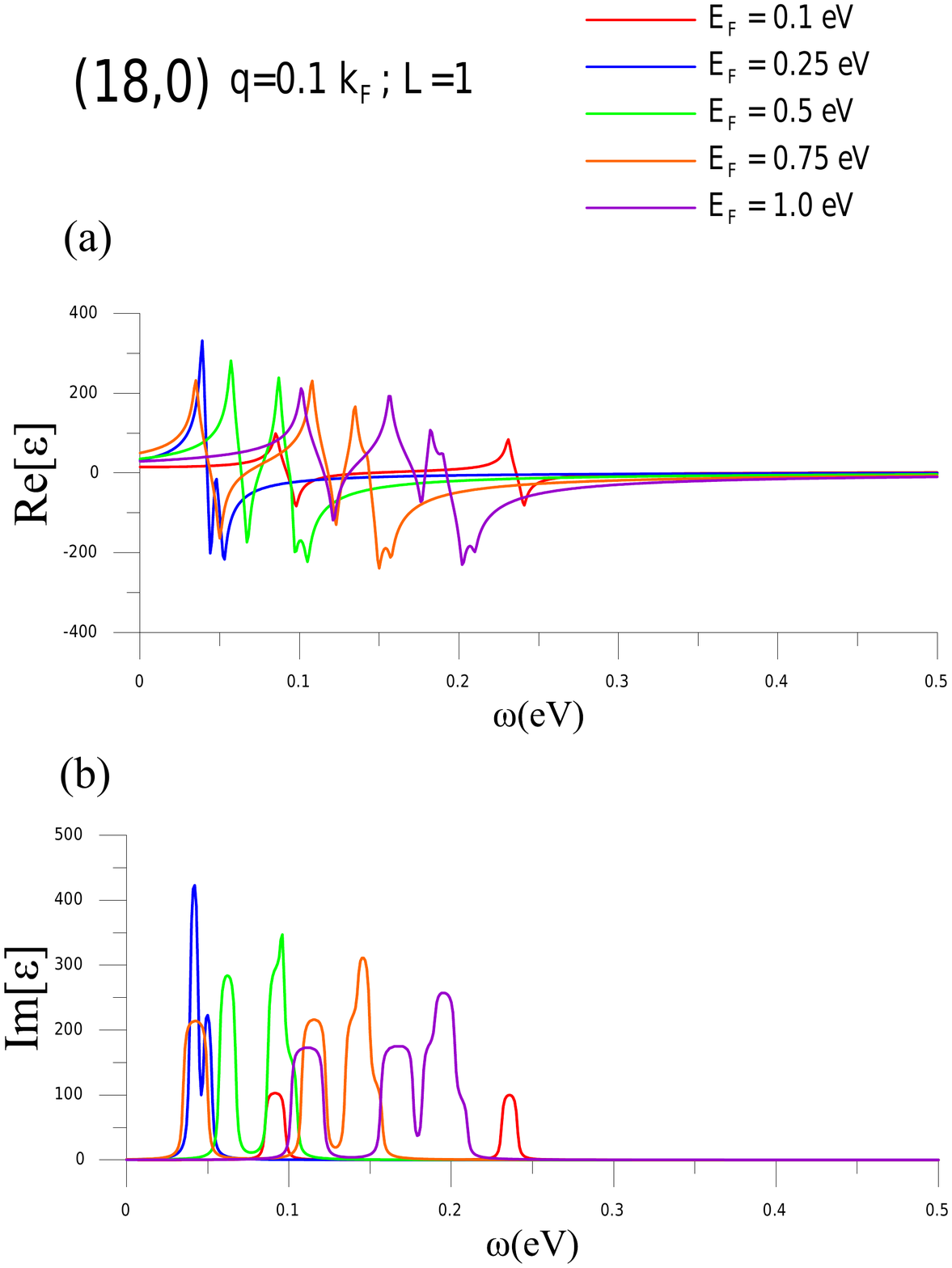}
\end{center}
\par
\textbf{fig10.9} The real-/imaginary-part dielectric function in a doped (18, 0) carbon nanotube for the ${L=0}$ mode (a)/(b) at ${q=0.1 k_F}$ and different Fermi levels, and (c)/(d) under ${E_F=0.5}$ eV and various  momentum transfers. The inset of (b) shows the inter-$pi$-band prominent response at the higher frequency beyond ${2E_F}$. The similar results of the ${L=1}$ mode are, respectively, illustrated, in (e)/(f) and (g)/(h).
\end{figure}

\newpage

\begin{figure}[tbp]
\par
\begin{center}
\leavevmode
\includegraphics[width=1.0\linewidth]{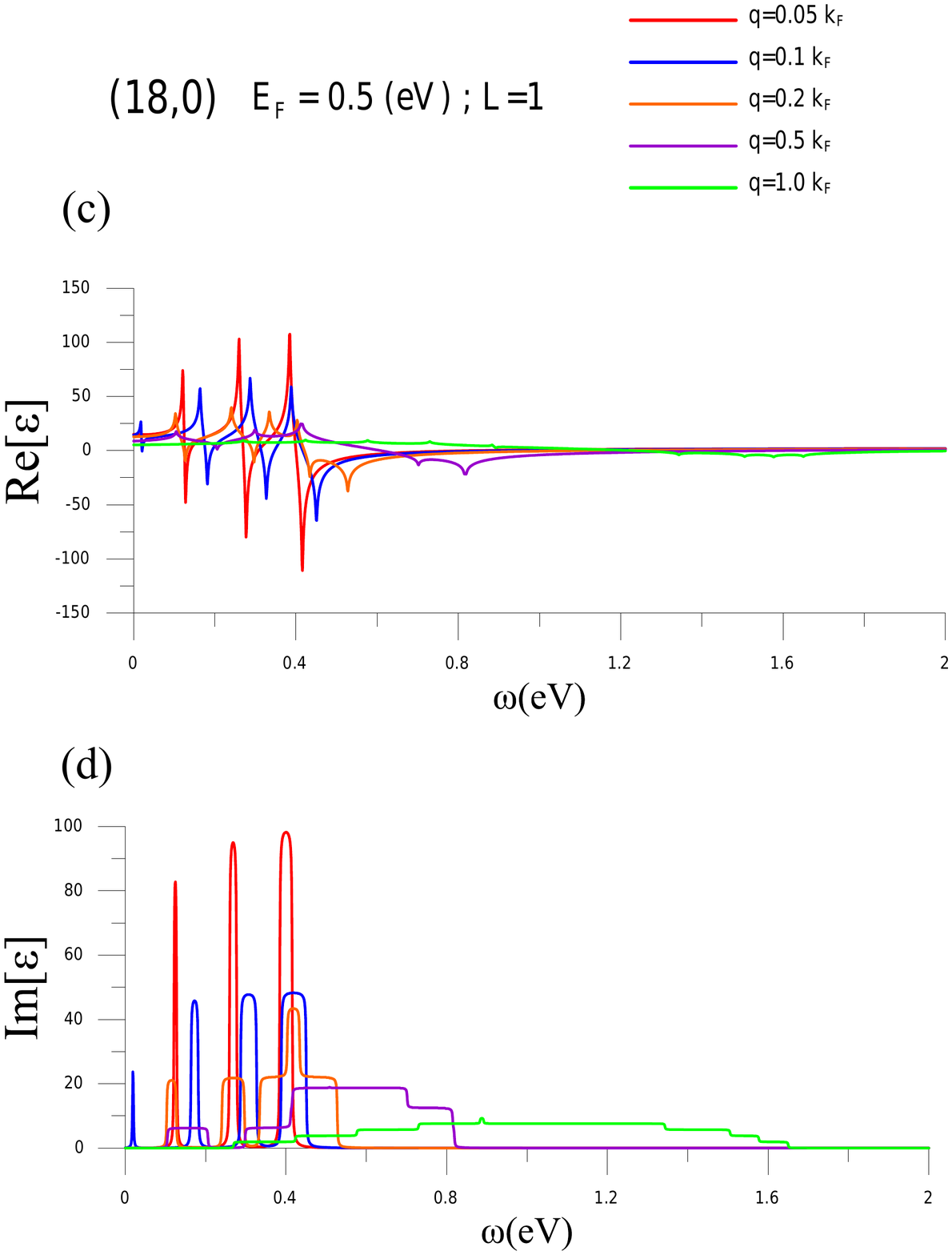}
\end{center}
\par
\textbf{fig10.9} The real-/imaginary-part dielectric function in a doped (18, 0) carbon nanotube for the ${L=0}$ mode (a)/(b) at ${q=0.1 k_F}$ and different Fermi levels, and (c)/(d) under ${E_F=0.5}$ eV and various  momentum transfers. The inset of (b) shows the inter-$pi$-band prominent response at the higher frequency beyond ${2E_F}$. The similar results of the ${L=1}$ mode are, respectively, illustrated, in (e)/(f) and (g)/(h).
\end{figure}

\newpage

\begin{figure}[tbp]
\par
\begin{center}
\leavevmode
\includegraphics[width=1.0\linewidth]{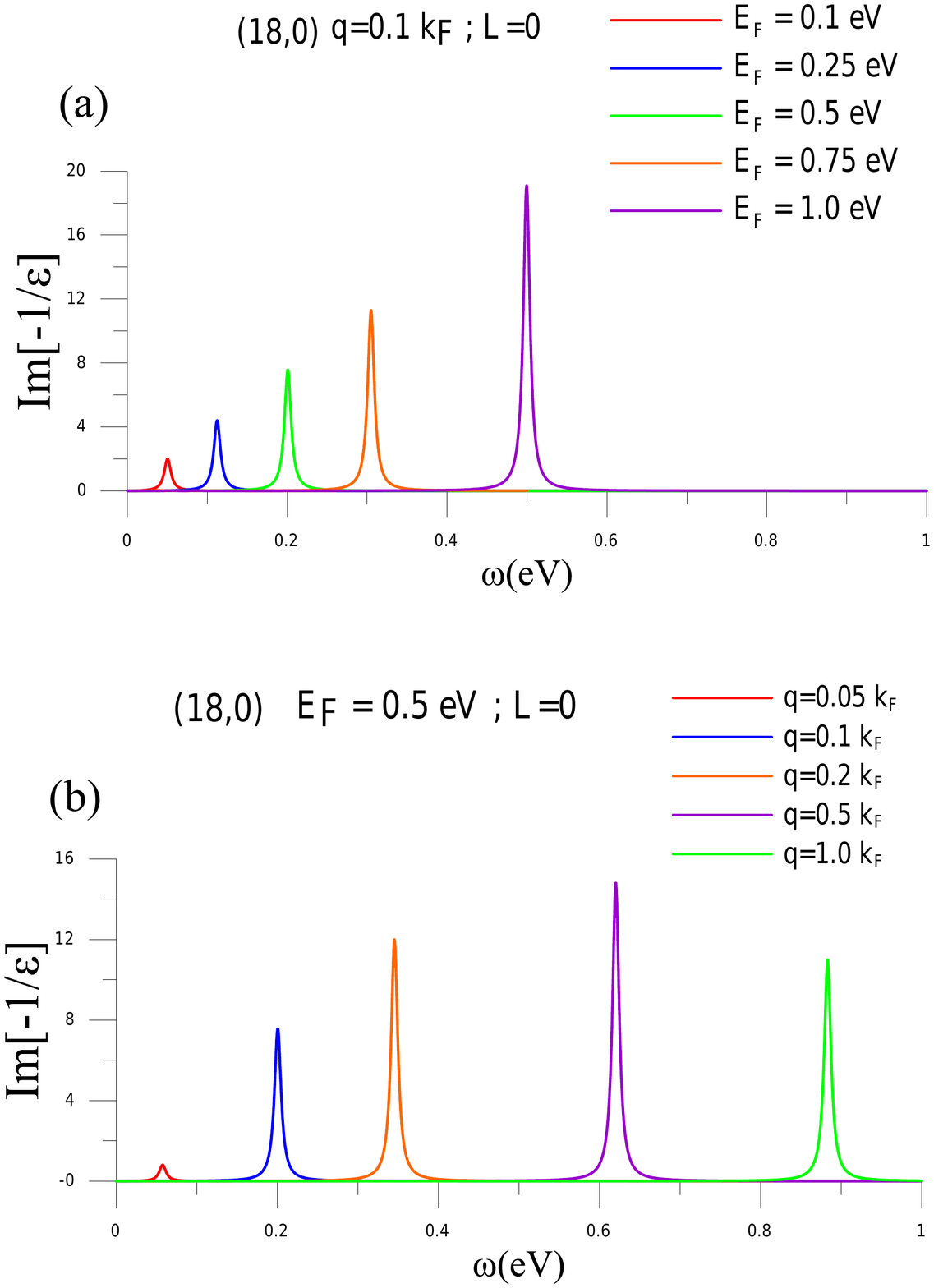}
\end{center}
\par
\textbf{fig10.10} The screened response functions of  $n$-type (18, 0) carbon nanotubes under the ${L=0}$ mode at (a) ${q=0.1}$ $k_F$ $\&$ distinct Fermi levels and (b) ${E_F=0.5}$ eV  $\&$ various transferred momenta. Also shown for the ${L=1}$ mode in (c) and (d).
\end{figure}

\newpage

\begin{figure}[tbp]
\par
\begin{center}
\leavevmode
\includegraphics[width=1.0\linewidth]{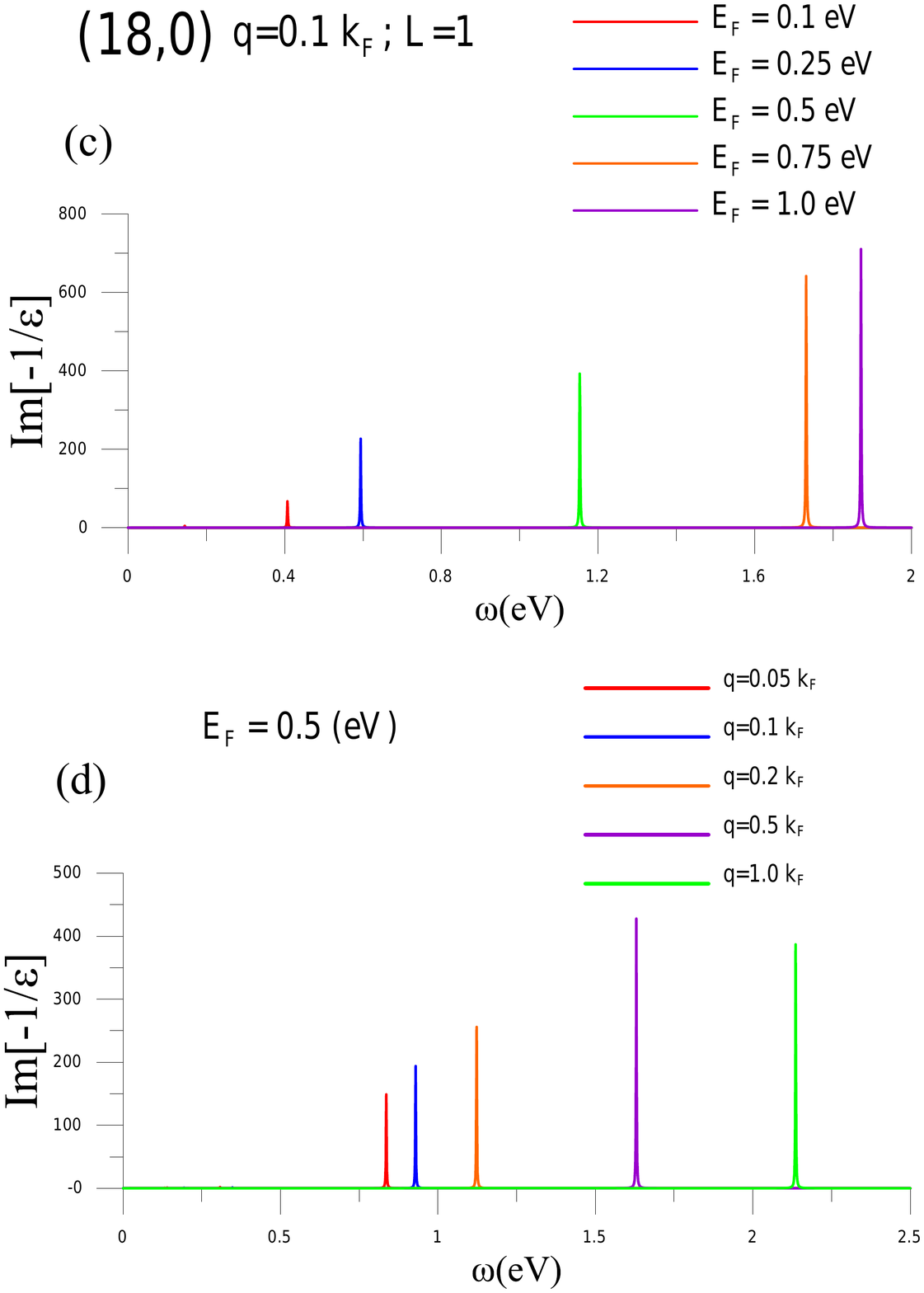}
\end{center}
\par
\textbf{fig10.10} The screened response functions of  $n$-type (18, 0) carbon nanotubes under the ${L=0}$ mode at (a) ${q=0.1}$ $k_F$ $\&$ distinct Fermi levels and (b) ${E_F=0.5}$ eV  $\&$ various transferred momenta. Also shown for the ${L=1}$ mode in (c) and (d).
\end{figure}

\newpage

\begin{figure}[tbp]
\par
\begin{center}
\leavevmode
\includegraphics[width=1.0\linewidth]{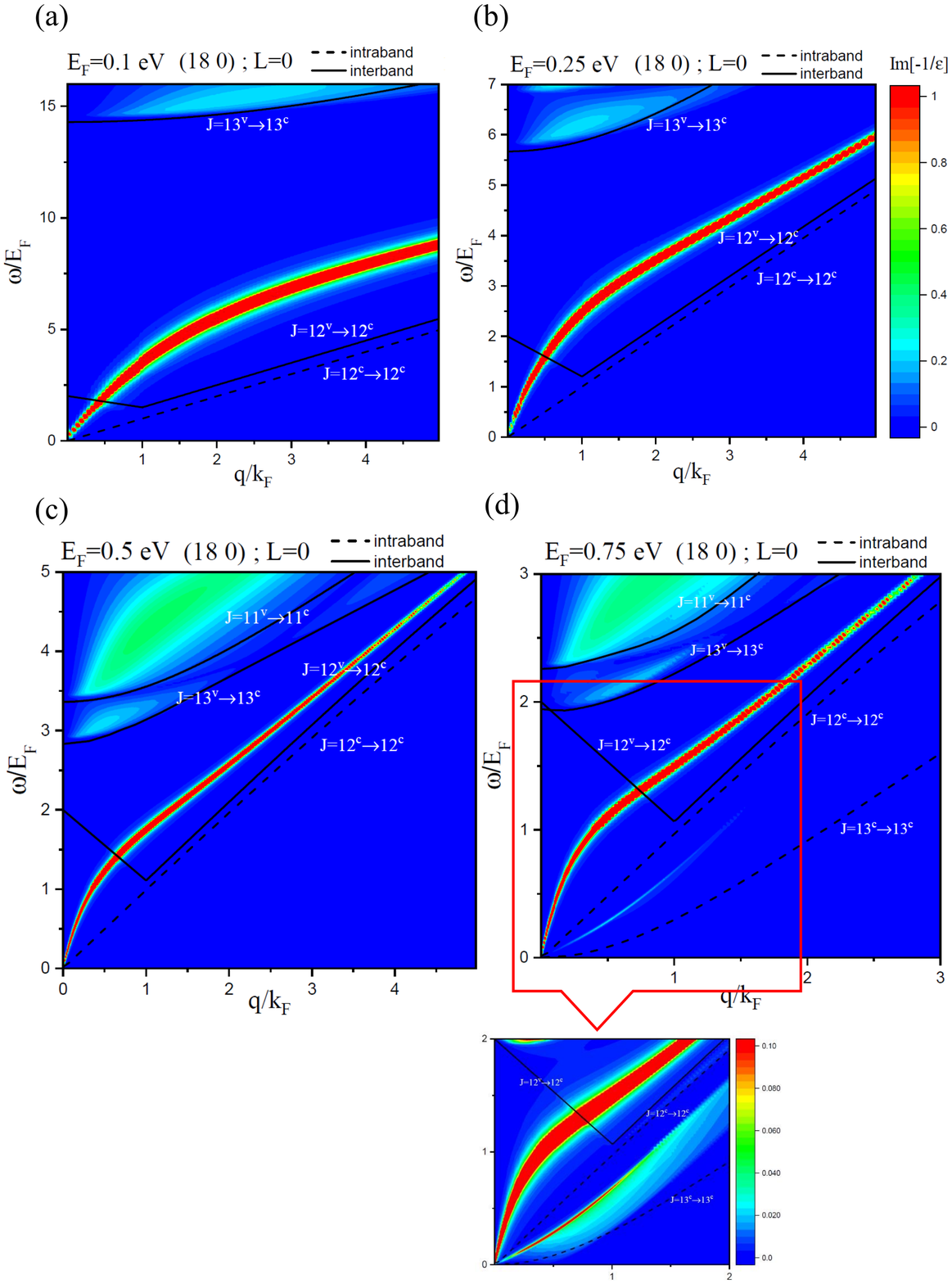}
\end{center}
\par
\textbf{fig10.11} The [${q, \omega}$]-excitation phase diagrams of metallic (18, 0) nanotubes under the specific ${L=0}$ mode with the different Fermi levels: ${E_F}$s=0.10, 0.25, 0.5, 0.75, 1.0 and 1.5 eVs, respectively, corresponding to (a), (b), (c), (d), (e) and (f).
\end{figure}

\newpage

\begin{figure}[tbp]
\par
\begin{center}
\leavevmode
\includegraphics[width=1.0\linewidth]{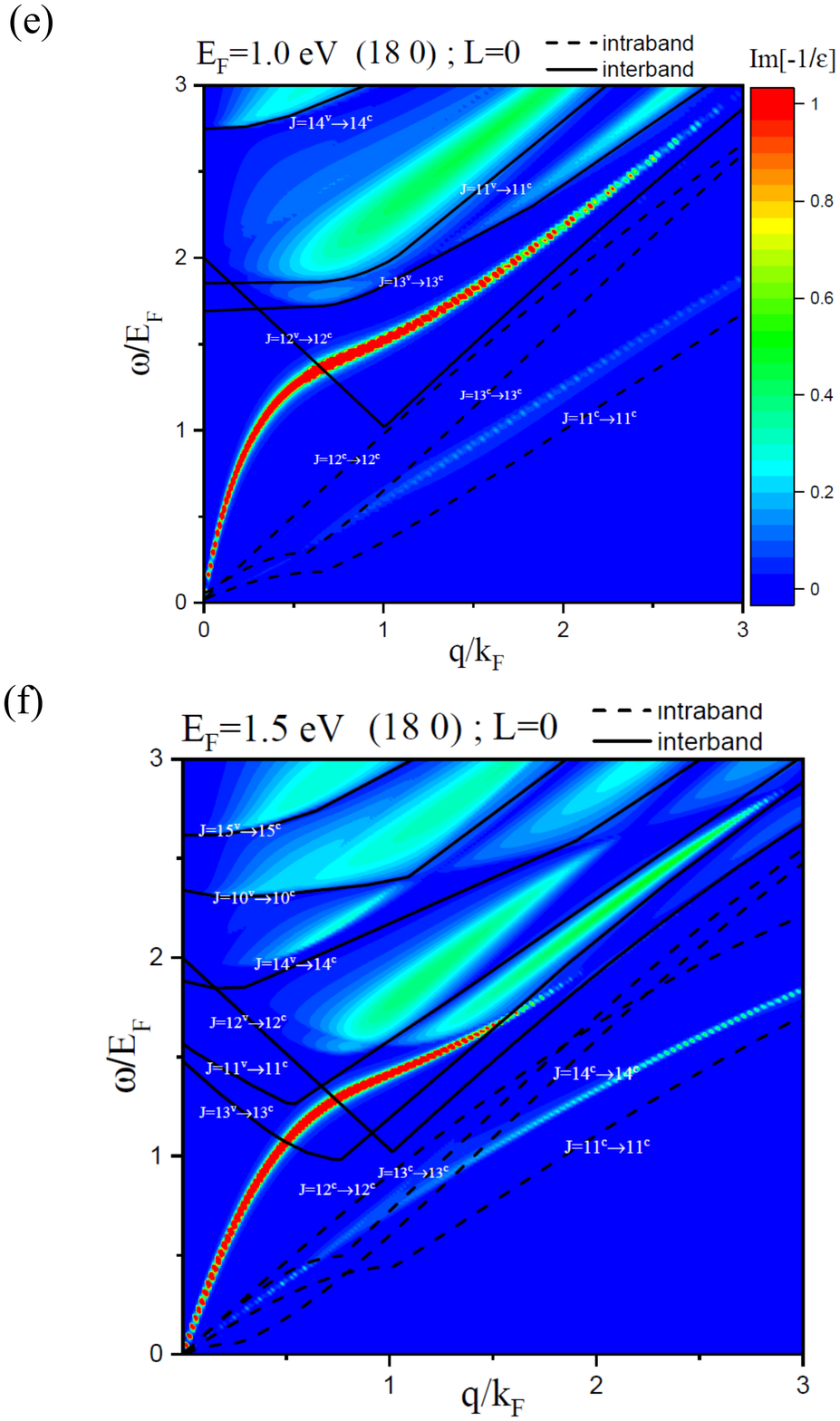}
\end{center}
\par
\textbf{fig10.11} The [${q, \omega}$]-excitation phase diagrams of metallic (18, 0) nanotubes under the specific ${L=0}$ mode with the different Fermi levels: ${E_F}$s=0.10, 0.25, 0.5, 0.75, 1.0 and 1.5 eVs, respectively, corresponding to (a), (b), (c), (d), (e) and (f).
\end{figure}

\end{document}